\documentstyle[namedreferences,psfig]{kluwer}

\let\oldverbatim\verbatim
\renewcommand{\verbatim}{\expandafter\small\oldverbatim}

\runningtitle{NEUTRON--INDUCED NUCLEOSYNTHESIS}
\runningauthor{H. OBERHUMMER ET AL.}

\begin{opening}
\title{Neutron--Induced Nucleosynthesis}

\author{H. \surname{Oberhummer}}
\author{H. \surname{Herndl}}
\institute{Institut f\"ur Kernphysik, Wiedner Hauptstr.~8--10, TU Wien,
A--1040 Vienna, Austria}
\author{T. \surname{Rauscher}}
\institute{Institut f\"ur theoretische Physik, Universit\"at Basel,
Klingelbergstr.~82, CH--4056 Basel, Switzerland}
\author{H. \surname{Beer}}
\institute{Forschungszentrum Karlsruhe, Institut f\"ur Kernphysik,
P.~O.~Box 3640, D--76021 Karlsruhe, Germany}

\date{}

\end{opening}

\begin{document}


\begin{abstract}
Neutron--induced nucleosynthesis plays an important role in
astrophysical scenarios like in primordial nucleosynthesis
in the early universe, in the
s--process occurring in Red Giants, and in the $\alpha$--rich
freeze--out and r--process
taking place in supernovae of type II\@. A review
of the three important aspects of neutron--induced nucleosynthesis is
given: astrophysical background, experimental
methods and theoretical models for determining reaction
cross sections and reaction rates at thermonuclear energies. Three specific examples
of neutron capture at thermal and thermonuclear energies
are discussed in some detail.
\end{abstract}

\keywords{Nucleosynthesis, s--process, r--process,
$\alpha$--rich freeze--out, primordial nucleosynthesis, neutron detection,
neutron activation, compound nucleus, direct reaction}

\section{Introduction}
For a long time it has been known that the solar--system abundances of elements
heavier than iron have been produced by
neutron--capture reactions~\cite{Burb57}. Neutron--induced
nucleosynthesis is also of relevance for abundances of isotopes
lighter than iron (especially for neutron--rich
isotopes), even though the bulk of these elements
has been synthesized by charged--particle induced reactions.

There are three main sites for nucleosynthesis: (i) primordial
nucleosynthesis forming the light elements H, He and Li in
the Big Bang, (ii) interstellar nucleosynthesis creating the
elements Li, Be and B and finally (iii) stellar nucleosynthesis
being responsible for the creation of all the elements from C to U.
In these astrophysical scenarios neutron--induced reactions
play a role in primordial nucleosynthesis
in the early universe, in the
s--process occurring in Red Giants, and in the $\alpha$--rich
freeze--out ($\alpha$--process) and r--process
thought to take place in supernovae of type II\@. In the early universe
the neutrons are formed through the hadronization of the quark--gluon
plasma. How\-ever, these neutrons decay soon due
to their mean lifetime of about 15 minutes. In Red Giants neutrons
are formed in helium burning through the reactions $^{22}$Ne($\alpha$,n)$^{25}$Mg and
$^{13}$C($\alpha$,n)$^{16}$O.

The above astrophysical sites and their relevance for neutron--induced
nucleosynthesis will be discussed in Sect.~2.
The experimental methods and measurements at thermal ($k_{\rm B}T = 25.3$\,meV)
and thermonuclear energies ($k_{\rm B}T \approx$ 1\,keV--1\,MeV)
are presented in Sect.~3. In that section, we focus
on neutron radiative capture, even though (n,p) and (n,$\alpha$) reactions
also play an important role in reaction networks up to $A \approx 40$.
For the experimental detection of the reaction
events mainly two signatures characterizing the capture process are discussed:
detection of promptly emitted capture $\gamma$--radiation (direct detection
methods) and the activity of the reaction product (activation methods).

A survey of the theoretical methods used in calculating
reaction cross sections and reaction rates and the underlying reaction
mechanisms, i.e., compound--nucleus (CN) formation and direct reaction (DI) 
mechanism is given
in Sect.~4. We discuss phenomenological methods (R--matrix, Breit--Wigner
formula) and the statistical model (Hauser--Feshbach) that are used
for calculations of the CN mechanism.
The methods used for investigating reactions in the DI are
microscopic methods (e.g.,\ Resonating Group Method and Generator
Coordinate Method) and potential models (Distorted Wave Born Approximation,
Direct Capture). Especially, potential models will be discussed in
some detail, because microscopic methods are limited to light nuclei
and rarely reproduce simultaneously experimentally known quantities
like separation energies,
resonance energies and low--energy scattering data.

Finally, some specific examples are given in Sect.~5. We start with
a detailed investigation of the reaction $^{36}$S(n,$\gamma$)$^{37}$S
at thermonuclear and thermal energies, since for this reaction the DI
mechanism is dominating. For the DI mechanism
detailed nuclear--structure information is necessary to calculate
cross sections. The comparison of the calculated capture cross
sections of neutron--rich targets using different nuclear--structure inputs with
the experimental data serves also as a benchmark for the calculation
of neutron--capture using nuclear--structure models to still
more neutron--rich isotopes off--stability.
In the reaction $^{208}$Pb(n,$\gamma$)$^{209}$Pb
the resonant CN--contribution at thermal and thermonuclear energies
is of the same order as the non--resonant DI contribution.
Taking this into account excellent agreement with the experimental data
is obtained. Finally, we investigate neutron--capture of Gd--isotopes at thermonuclear
energies as a specific example where only CN contributions are of relevance.
For these reactions
statistical Hauser--Feshbach calculations for the cross sections at
thermonuclear energies are compared with recently measured
experimental data.

The paper is concluded by an appendix for the reader looking for more
details, giving important formulas and definitions which
were too specific to be included in the main body of the text.

\section{Astrophysical scenarios}
Nucleosynthesis theory predicts that the formation of most of the nuclear
species of mass with about $A>60$ occurs as a consequence of neutron--induced
reactions.
Charged--particle thermonuclear reactions dominate the production of the
heavy elements through approximately iron and nickel. Beyond the iron group,
however, neutron--capture processes are favored because of the increasing
Coulomb barriers which can only be overcome at such high temperatures, so that
photodisintegration hinders the build--up of heavy nuclei via charged particle induced
reactions.
In the past years, a number of neutron--rich astrophysical environments
have been suggested
to explain the abundance pattern of heavy nuclei,
in all of which neutron--induced reactions play an important role.
Among these are the suggested sites for the s-- and r--processes, but also more
exotic scenarios such as primordial nucleosynthesis in inhomogeneous
cosmologies. In this chapter we will briefly discuss the mentioned
processes of nucleosynthesis, starting with the s-- and r--processes
and their respective sites. Additional information can be found also in the
reviews of
\citeauthor{Arnould90}~\shortcite{Arnould90},
and \citeauthor{Mey94}~\shortcite{Mey94}.

When examining the heavy--element abundances of solar system matter~\cite{And89,pal93}
one can find features correlated with the positions of the neutron shell
closures at $N=50$, 82, and 126. The splitting of those abundance peaks (in the
regions $A=$80--90, 130--140, 190--210) suggests at least two scenarios with
very distinct neutron fluxes.
Historically, this has led to the definition of two main
nucleosynthesis processes, namely the s-- (slow neutron capture) and the
r--process (rapid neutron capture), which are thought to take place in
quite different astrophysical environments. Common to both is that some
pre--existing distribution of seed abundances is exposed to
neutron--irradiation,
by which heavier elements are produced through subsequent neutron--captures
and beta--decays. The distinction is made by comparison of the relative
lifetimes for neutron captures ($\tau_{\rm n}$) and beta--decays
($\tau_{\beta}$). In the case of the s--process we find
$\tau_{\rm n}>\tau_{\beta}$, for the r--process it is
$\tau_{\rm n}\ll \tau_{\beta}$. (The definitions of the lifetime
$\tau_{\rm n}$ and the related astrophysical reaction rates can be
found in Appendix \ref{astroapp}).

Although the temperatures encountered in astrophysical environments are
often of the order of billions of Kelvin, the corresponding energies at which
the nuclear cross sections have to be known are relatively low by nuclear
physics standards. For neutron--induced reactions one can derive a simple
formula for the relevant energy range. The velocity distribution of the
interacting particles follow the Maxwell-Boltzmann distribution at a given
temperature. The dominant peak of that distribution (i.e., the (by far) most
likely energy) is found at $E=k_{\rm B}T$. When using convenient units, this
leads to $E=0.0862\,T_9$, with $E$ measured in MeV
(1\,eV = $1.60219 \times 10^{-19}$\,J) and $T_9$ denoting the
temperature in $10^9$\,K. The maximum temperatures in nucleosynthesis
processes are $T_9 =$7--10. Thus, the energies are limited to the
range of hundreds of keV and below.

\subsection{The s--Process}
The condition $\tau_{\rm n}>\tau_{\beta}$ ensures that the neutron--capture
path will remain close to the valley of beta--stability. If an unstable
nucleus is encountered in the neutron--capture chain, usually beta--decay
to the next higher element is much faster than another neutron capture.
Therefore, the resulting abundances are determined by the respective
neutron--capture cross sections. Isotopes with small cross sections act as
bottlenecks and build up large abundances. This leads to s--process
peaks at magic neutron numbers, where the cross sections are particularly
small. These bottleneck isotopes represent neutron exposure monitors determining
the required mean s--process exposure. Accurate neutron--capture cross sections
are also needed
to study the properties of the s--process
branchings in the capture chain where the rates for capture and beta--decay
are comparable. In these branchings very often a group
of two, three or four radioactive isotopes is involved (e.g., $^{151}$Sm,
$^{152,154}$Eu, and $^{153}$Gd). Investigation of the abundance patterns
in such branchings
can yield a variety of important information, such as estimates of the
neutron density, temperature, and electron density
during the synthesis.
It turned out that three different s--process components
are necessary for a satisfactory description of the
observed s--process abundances: The bulk of s--process material in the
mass range $90 \le A \le 204$ is produced by what is called the {\em main}
s--process component, an s--process with an exposure distribution $\rho(\tau)$
which is smoothly decreasing for increasing neutron exposures
$\tau$~\cite{Seeger65,gal96}.
Especially an exponential exposure distribution is chosen which is considered to
be the result of a pulsed neutron flux \cite{ulrich73}.
A {\em weak} component --- characterized by a smaller but continuous
mean neutron exposure --- is added to describe the abundances below
$A\simeq 90$. Finally, a {\em strong} component is to be postulated to
account for the abundance maximum at lead. For more detailed
reviews on the s--process see~\citeauthor{Kaepp89}~\shortcite{Kaepp89},
and \citeauthor{Mey94}~\shortcite{Mey94}
and references therein.

The analyses of s--processing with parametrized models --- which can be
basically undertaken independently of the true astrophysical sites and neutron
sources --- had proven to be quite successful. The s--process calculations
can be found in \citeauthor{Kaepp89} \shortcite{Kaepp89}, \citeauthor{beer91}
\shortcite{beer91}, and \citeauthor{pal93} \shortcite{pal93}. In a recent
development of the main component with a parametrized model using combined
burning of two neutron sources (\citeauthor{bco96}, \citeyear{bco96}, \citeyear{bcm96}),
new clues have been found to the Pb--Bi formation at the s--process termination.
One important outcome of these investigations is the empirical
r--process abundance distribution obtained
by subtraction of the calculated s--process abundances from the solar
abundances (\citeauthor{Kaepp89}, \citeyear{Kaepp89}; \citeauthor{pal93},
\citeyear{pal93}; \citeauthor{bco96}, \citeyear{bco96}, \citeyear{bcm96}).
The search for the astrophysical s--process site(s) is still
somewhat controversial~\cite{Arnould90}. As neutron sources,
the most promising reactions are $^{22}$Ne($\alpha$,n)$^{25}$Mg and
$^{13}$C($\alpha$,n)$^{16}$O~\cite{Cameron55,Burb57,Reeves66,Kaepp94}. Both reactions
are typical of He--burning environments. Large amounts of $^{22}$Ne can be
expected to be built up at the very beginning of the He--burning phase.
However, the $^{22}$Ne neutron source requires rather high temperatures
($T\geq 3 \times 10^8$\,K) to be activated. On the other hand, the $^{13}$C
reaction can easily take place at lower temperatures (radiative burning at
$T \simeq 0.9 \times 10^8$\,K, convective burning at $T \simeq 1.5
\times 10^8$\,K) but the sufficient supply of $^{13}$C poses a problem. In
current models both neutron sources are activated successively
\cite{Kaepp90,stran95,gal96}.

A long series of research has been dedicated to the analyses of the
He--burning phases in stars of different masses.
It appears that the s--process components found in the classical
model can be attributed to different sites. So it is now commonly believed
that the weak component has to be ascribed to core He--burning in massive
stars ($M \geq 15\,M_{\odot}$, 1 M$_{\odot}$ = $1.989\times 10^{30}$\,kg),
but C and Ne burning in the outer layers of the
star (``shell burning'') may also be important.
He--shell burning in intermediate and
low mass stars ($M < 8 M_{\odot}$) could supply the conditions required for
the main component~\cite{Kaepp90,stran95}. The origin of the strong component
is less understood but it was suggested that the core He--flash in stars
of less than 11\,$M_{\odot}$ could be responsible for that contribution.
But there is probably no need for a separate site if the distribution of
exposures in the main component is not exactly exponential but is higher than
exponential at large exposures $\tau$~\cite{Mey94}.

\subsection{The r--Process}
Approximately half of all stable nuclei observed in nature in the heavy
element region about $A>60$ is produced in the r--process. This r--process occurs
in environments with large neutron densities which lead to
$\tau_{\rm n} \ll \tau_{\beta}$. Contrary to the s--process, the successive
neutron captures will proceed into the region of neutron--rich nuclei far--off
stability. For the large neutron fluxes characteristic of this process,
the closed neutron shells are encountered in the neutron--rich regions and thus at
lower proton numbers. Therefore, the yield peaks are located
at slightly lower mass number after the beta--decay
of the products to stability, than in the s--process. In contrast to the
beta--decay lifetimes
of critical nuclei participating in the s--process in the range from about
10 to 100 years, the most neutron--rich isotopes along the r--process
path have lifetimes of less than one second and more typically 10$^{-2}$
to 10$^{-1}$\,s. Cross sections for most of the nuclei involved cannot
be experimentally measured anymore due to the short half--lives. Therefore,
theoretical descriptions of the capture cross sections as well
as the beta--decay half--lives are the only source of the nuclear physics
input for r--process calculations. For nuclei with about $Z>80$ beta--delayed
fission and neutron--induced fission might also become important.

For realistic, dynamic r--process calculations following temperatures
and neutron densities changing rapidly with time (such as in investigations
of freeze--out effects), it proves necessary to use a complete reaction
network with all relevant reaction and decay rates included. However,
important information about the required r--process temperatures and
densities can be derived from parameter studies using simplified models.
At high neutron number densities ($>10^{20}$\,cm$^{-3}$) and high temperatures
(i.e., large high--energy photon density, $T>10^9$\,K) photodisintegrations
will be active and balance the capture flow. Beta--decay lifetimes are
usually longer than (n,$\gamma$)-- and ($\gamma$,n)--time scales and thus
each isotopic chain for any $Z$ will be populated by an equilibrium
abundance. In this case, the differential equations governing the
reaction network can be simplified in such a way that the resulting
abundance ratios are only dependent on the neutron density, temperature
and the neutron separation energies~\cite{Cowan91}.
This is the so--called {\em waiting--point
approximation} because the nucleus with maximum abundance in each isotopic
chain must wait for the longer beta--decay time scale. In such an
(n,$\gamma$)$\rightleftharpoons$($\gamma$,n) equilibrium, no detailed
knowledge of neutron--capture cross sections is needed.

Another simplified approach is the {\em steady--flow approximation}.
Again, the treatment of the full network can be facilitated by assuming
an equal flux via beta--decays into an isotopic chain with charge number
$Z$ and out of the chain $Z$ into $Z+1$~\cite{Cowan91}.
After a time larger than the longest beta--decay half--lives, and if fission cycling
(see below)
is neglected, all the nuclei in the network will approach such
steady--state abundances. This simplifies the solution of the
equations because the assumption of an abundance at one $Z$ is
sufficient to predict the whole r--process curve. However, although steady--flow
calculations correctly treat the neutron--capture rates and beta--decays
of individual $Z$--chains, the effects of neutron captures and beta--decays
on the dynamics of the r--process are ignored. Depending on the time scales involved,
this may or may not be a valid simplification. Therefore, this method
is useful when studying a number of different r--process conditions
but it is not always comparable to a full dynamic network calculation.

In the quest for finding the site of the r--process, parameter studies
employing the approximations described above yielded important
information to put con\-straints on the required conditions. Additionally,
they could also show deficiencies in our knowledge of nuclear physics
regarding theoretical mass models and thereby spur a series of
investigations aimed at improving the nuclear physics input~\cite{Thiele94}.
Making use of
(n,$\gamma$)$\rightleftharpoons$($\gamma$,n) equilibrium,
a recent analysis~\cite{Thiele93,Kratz93} of the
solar system isotopic r--process abundance pattern revealed
that it can only be reproduced with (at least) three components
with different neutron densities ($>10^{20}$\,cm$^{-3}$). This is
necessary for correct positions of
the three abundance peaks at $A=80$, 130, and 195. These
three components
(at time scales of about 1.5--2.0\,s) also establish a steady flow
for beta--decays in between magic neutron numbers. The steady flow
breaks down only at magic neutron numbers where the r--process path comes
closest to
stability and encounters the longest beta--decay
half--lives.
A local steady flow behavior had been proven with dominantly
experimental mass and half--life knowledge at
the magic neutron numbers N=50 and 82  for $^{77}_{27}$Co to $^{80}_{30}$Zn
as well as  $^{127}_{\phantom{1}45}$Rh to $^{130}_{\phantom{1}48}$Cd
\cite{Kratz88}. Here the path comes closest to stability and
experiences increasingly long beta--decay half--lives, before the steady
flow breaks down beyond the abundance peaks with the longest half--lives
(which must therefore be comparable with the process time itself). It is
then obvious that a steady flow has to apply for the shorter half--lives
in between closed shells. The propagation of an
r--process will follow a contour line for a specific neutron separation
energy
$S_{\rm n}$~\cite{Thiele94}. From deviations of the calculated compared to
the observed abundances one can then draw conclusions on the
validity of the nuclear theory used at a certain $S_{\rm n}$.

Concerns that this might be an overinterpretation have been raised since
\cite{Arnould93,Howard93}, and
that an almost continuous superposition of a multitude of
components would automatically be able to prevent the deviations
from the solar abundance pattern attributed to deficiencies in nuclear
mass models. However, it was shown that this is not
possible \cite{Kratz94,Chen95,vero96} and that therefore there is still an
urgent need for improved nuclear theory in astrophysics.

Despite many efforts, the site of the r--process has not been clearly
identified yet, although there are strong clues from recent studies
and observations. The basic requirements for the r--process are neutron
number densities $n_{\rm n}>10^{20}$\,cm$^{-3}$ and temperatures around
$10^9$\,K. In order to be able to synthesize the heaviest elements at
$A\approx 240$ one also needs a sufficient supply of neutrons (180 per
seed nucleus, when starting at $A\approx 60$).
An average over all isotopes produced gives a mean value of
80 neutrons per seed nucleus. Thus, we arrive at another constraint
for the abundance ratio of neutrons over seed nucleus of
$1<Y_{\rm n}/Y_{\rm seed}<180$.

The short time scale (i.e., seconds) for neutron capture and the
correspondingly large neutron fluxes required have for some time
suggested an explosive astrophysical origin of the r--process.
The currently most favored environment is
the high--entropy bubble of an exploding type II
supernova. The iron core of a typically 8--25\,M$_{\odot}$ star will
collapse after Si--burning when exceeding the Chandrasekhar mass limit.
The core will become stable again at nuclear density when the nucleon--gas
becomes degenerate.
The core will bounce back
and a shockwave will run through the outer layers of the collapsed star.
However, this shockwave does not carry enough energy to explode the star.
The shock is heating the material to such high temperatures that the
previously produced iron is photodisintegrated again. This process takes
4--7\,MeV per nucleon (7\,MeV for a complete photodisintegration into nucleons)
and will eventually halt the shockfront.
The formation of the neutron star leads to a gain in gravitational binding
energy which is released in the form of
neutrinos. Although the interaction of neutrinos with matter is quite
weak, considerable amounts of energy can reheat the outer layers even
if only 1\% of the 10$^{53}$\,erg (10$^{46}$\,J) in neutrinos
is deposited via neutrino captures on neutrons and protons.
This accelerates the shockwave
again and can finally explode the star. Because of the heating and
expansion of the gas, a zone with low density and high temperature will
be formed behind the shockfront, the so--called {\em high--entropy bubble}
\cite{Herant94,Burrows95,Janka95}.

The nucleosynthesis in the high--entropy bubble is thought to proceed as
follows. Due to the high temperature, the previously produced nuclei up
to iron will be destroyed again by photodisintegration. At temperatures
of about 10$^{10}$\,K the nuclei would be dismantled into their
constituents, protons and neutrons. At slightly lower temperatures one is
still left with $\alpha$--particles. During the subsequent cooling of
the plasma the nucleons will recombine again, first to $\alpha$--particles,
then to heavier nuclei, starting with the reactions
3$\alpha \rightarrow ^{12}$C and $\alpha+\alpha+{\rm n} \rightarrow ^9$Be, followed
by $^9$Be($\alpha$,n)$^{12}$C.
Depending on the exact temperatures, densities
and the neutron excess, quite different abundance distributions can be
produced in this {\em $\alpha$--rich freeze--out} (sometimes also called
{\em $\alpha$--process}, not to be confused with the $\alpha$--process
erronously defined by \citeauthor{Burb57}, \citeyear{Burb57}), as compared to the nuclear
statistical equilibrium found in the late evolution phases of massive
stars~\cite{Arnett71,Woos73,Woos92,Frei95}. Temperature and density are
dropping quickly in the adiabatically expanding high--entropy bubble.
This will hinder the recombination of alpha particles into heavy nuclei,
leaving
a high $Y_n/Y_{\rm seed}$ and sufficient neutrons for an r--process,
(acting on the newly produced material)
at the end of the $\alpha$--process after freeze--out of charged particle reactions.

It should be noted that there are still many open questions and that we
still lack a complete, quantitative understanding of the explosion
mechanism of type II supernovae, although progress has been made with
improved neutrino transport schemes~\cite{Wilson93,Janka94} and in
multidimensional calculations
(\citeauthor{Burr92}, \citeyear{Burr92};~\citeauthor{Burrows95},
\citeyear{Burrows95};
\citeauthor{Janka93}, \citeyear{Janka93}; \citeauthor{Janka96},
\citeyear{Janka96};
\citeauthor{Hera92}, \citeyear{Hera92}, \citeyear{Herant94}).
Therefore, consistent calculations like
\citeauthor{Woosley94}~\shortcite{Woosley94} or \citeauthor{Takaha94}~\shortcite{Takaha94}
still bear some uncertainties, especially with respect to the entropies actually
obtained. Only a mass of 10$^{-6}$ to 10$^{-4}$\,M$_{\odot}$
of processed matter has to be ejected per supernova event to provide
the quantities derived from galactic properties~\cite{Truran71}.

Due to the large difficulties still encountered in type II supernova
calculations, a variety of other models for possible r--process sites
had been suggested. Common to all these scenarios is, of course, that
they have to be able to provide the necessary conditions for the
r--process, as described above. Among these are ``bubbles'' or ``jets''
ejected by the collapse of rotating stellar cores, accretion discs
around colliding neutron stars (neutron star mergers) and
collisions between a neutron star and a black hole
(see~\citeauthor{Cowan91} \citeyear{Cowan91}
and references therein). The possibility of
a primordial production of at least a floor of r--process elements is
discussed below.

\subsection{Primordial Nucleosynthesis}
According to the big bang model of cosmology the early Universe
could provide the conditions for nucleosynthesis processes creating
light elements up to lithium. For more detailed reviews on this topic
see, e.g., \citeauthor{Schramm77}~\shortcite{Schramm77},
\citeauthor{Boesgard85}~\shortcite{Boesgard85},
\citeauthor{Schramm95}~\shortcite{Schramm95}.
The synthesis does not immediately start
at weak freeze--out ($T\approx 1$\,MeV, age of the Universe $\approx 1$\,s)
because of the large number of photons relative to nucleons
($\eta^{-1}=n_{\gamma}/
n_{\rm b}\approx 10^{10}$). As the nucleosynthesis chain begins with the
formation of deuterium through the process p+n$\rightarrow$D+$\gamma$, it
is delayed past the point where the temperature has fallen below the
deuterium binding energy, which is at $T\approx 0.1$\,MeV. Nucleosynthesis
proceeds by further neutron, proton and light nuclei capture on deuterium,
to form $^3$H, $^3$He, $^4$He (which is the dominant product of big bang
nucleosynthesis with an abundance close to 25\,\% by mass), and even
heavier nuclei. However, the gaps existing among stable nuclei at mass
numbers $A=5$ and $A=8$ inhibit the formation of nuclei beyond $A=8$.
Therefore, the standard big bang can produce only D, $^3$He, $^4$He, and
$^7$Li in appreciable amounts. The strength of the standard big bang scenario
is that only one free parameter ($\eta$) must be specified to determine
all the primordial abundances, ranging over 10 orders of
magnitude~\cite{Olive90,Walker91}.
The excellent agreement between observed and predicted abundances forms
one of the cornerstones supporting the big bang model.

A number of possible mechanisms have been suggested
to generate density inhomogeneities in the early Universe
which could survive until the onset of primordial
nucleosynthesis~\cite{Mala93}. Such inhomogeneities could change primordial
nucleosynthesis in such a way as to enable the production of heavier
elements
by by--passing the mass gaps through the reaction sequence\\
$^7$Li(n,$\gamma$)$^8$Li($\alpha$,n)$^{11}$B(n,$\gamma$)$^{12}$B($\beta^-
$)$^{12}$C(n,$\gamma$)$^{13}$C(n,$\gamma$)$^{14}$C\dots\\
This sequence is inhibited in the relatively proton rich standard
big bang nucleosynthesis. However, in inhomogeneous models the ``bubbles''
of density fluctuations translate into regions of different
neutron--to--proton ratios, due to the different mean free diffusion paths
of the neutral neutrons and the electrically charged protons. Thus,
it becomes possible that neutrons are even over--abundant in the low
density zones, whereas the protons remain trapped in the high--density
regions. While in a proton--rich region charged--particle reactions will
play the dominant role (just as in the standard big bang),
neutron--induced reactions will be important in the neutron--rich
environment~\cite{Apple85,Sale86,Apple88}.

It was found that not only elements up to carbon and oxygen could be
produced but that it was even possible to synthesize r--process nuclei.
Although this is done in the low density regions, large abundances can
be obtained by {\em fission cycling}~\cite{Seeger65}. In an r--process with
fission cycling the production of heavy nuclei is not limited
to the r--process flow coming from
light
nuclei but requires only a small amount of fissionable nuclei to be
produced
initially. The total mass fraction of heavy nuclei is doubled with each
fission cycle
and can thus be written as $X_{\rm r}= 2^n X_{\rm seed}$,
with $X_{\rm seed}$ denoting the
initial
mass fraction of heavy nuclei.
The cycle number $n$ is decreasing with decreasing neutron number
density $n_{\rm n}$ (and increasing temperature $T$) because the
reaction flux experiences longer half--lives when the r--process path is
moving closer to stability.

Because of the exponential increase in r--process abundances, with a
sufficient number of cycles during the primordial nucleosynthesis era one can
even arrive at abundances exceeding the ones found in the solar system.
Thus, nucleosynthesis can put severe constraints on the conditions found
in the early Universe. Contrary to previous estimates, though, it was
found that r--processing will only occur at conditions already ruled out
by the light element abundances found in the proton--rich, ``standard''
zones~\cite{Rauscher94}. Although inhomogeneous
big bang nucleosynthesis could not fulfill the hopes put in it initially
(e.g., providing the means of setting $\Omega_{\rm b}=1$ in accordance
with the inflationary model), it still has to be considered since one
still cannot completely rule out the existence of density
inhomogeneities in the early Universe. (Note, however, that these
inhomogeneities would be on a scale far too small to solve the current
problems in explaining galaxy formation and the large--scale structure
of the Universe. The characteristic length of such a region has expanded to
a current value of 10$^{14}$\,m now~\cite{Rauscher94}. This is about
1000 times the distance
from Earth to the Sun and about 1\% the distance to the nearest star,
tiny by astronomical standards.)

\section{Experimental Methods and Measurements}
Although (n,p) and (n,$\alpha$) reactions play an important role in
the reaction networks of light isotopes up to the calcium isotopes
we will focus our discussion on experimental methods for
the detection of the neutron radiative capture
process that occurs at all mass numbers. In some cases
the (n,p) and (n,$\alpha$) reactions even dominate
over the (n,$\gamma$) process and hence modify the
nucleosynthesis path (see, e.g., \citeauthor{Scha95} \citeyear{Scha95};
\citeauthor{Wag95} \citeyear{Wag95}).

The measurement of neutron radiative capture reaction cross sections for
astrophysics
requires neutron sources covering an energy range from a few eV to 500\,keV
and detection telescopes for the counting of the reaction
events. For the production of a neutron flux of a sufficient strength,
re\-search reactors, but especially accelerators (e.g., Van de Graaff and
electron linear accelerators), are currently used.
For the detection of the reaction events, signatures
characterizing the capture process are applied: the promptly emitted
capture $\gamma$--radiation and the activity of the reaction product. The
first method, which we will call the {\em direct detection method,}
can be applied in principle to each isotope,
whereas the second method (decay counting) requires an unstable
reaction product with suitable decay characteristics (e.g.,
convenient half--life, strong $\gamma$--ray decay lines).
The detection of stable or long lived--reaction products by atom
counting methods is not yet fully developed and will
not be discussed here.

\subsection{Direct Detection Methods}
The neutron capture process on an isotope $^AZ$
leads to a final nucleus and $\gamma$--radiation:
$^AZ$ + n $\rightarrow$ $^{A+1}Z$ + $\gamma$.
The reaction energy $E^*$ consists of
the kinetic energy of the neutron
$E_{\rm kin}$ plus the neutron binding energy $E_{\rm b}$:
\begin{eqnarray}
E^*=E_{\rm kin}\frac{A}{A+1} + E_{\rm b} \quad .
\end{eqnarray}
The promptly emitted $\gamma$--radiation that accompanies the capture process
carries away the reaction energy $E^*$ and
consists in general of $\gamma$--ray cascades to the ground state with
varying multiplicity.

\subsubsection{High Resolution Capture $\gamma$--Ray Detection}
In a number of cases, especially in light nuclides
with well--known level schemes up to the region of excitation, it is
possible to measure individually all primary $\gamma$--ray transitions
with a detector
of good $\gamma$--energy resolution and
to integrate partial cross sections directly to obtain the total
capture. At the 3.2\,MV Pelletron Accelerator of the Research Laboratory for
Nuclear Reactors, Tokyo Institute of Technology this method~\cite{iga94}
has been applied
to determine the important stellar reaction rates of p(n,$\gamma$)d \cite{suz95},
$^{7}$Li(n,$\gamma$) \cite{nai91}, $^{12}$C(n,$\gamma$) \cite{nag91,ots94},
and $^{16}$O(n,$\gamma$) \cite{nag95}. The capture $\gamma$--rays were detected by an
anti--Compton NaI(Tl) detector and the neutron energy was determined by
time--of--flight.

It has been shown~\cite{coce94} that for heavier isotopes with more
complicated spectra
the capture rate can be obtained also by summing up all transitions ending at the
ground state. The obvious advantage as compared to the sum over primary
$\gamma$--rays is that low--energy transitions are usually stronger, better
resolved and detected with higher efficiency than the primary ones.

\subsubsection{Total Absorption Detection}
In principle the most straightforward method to detect capture events
independent from the details of the prompt $\gamma$--ray cascades is to
sum over all $\gamma$--cascades to obtain a signal proportional to
$E^*$=$E_{\gamma}^{\rm tot}$=$\sum_i E_{\gamma i}$. An ideal detector
covering the entire solid angle of 4$\pi$
would then yield a spectrum of capture events consisting of a peak at the energy
$E^*$. However, in differential detection systems (using
the time--of flight--method) is is difficult to separate signals
from scattered neutrons and true capture events. Therefore,
the detector should be insensitive to scattered neutrons as scattering
is about 10 times more likely than capture in the energy region of
astrophysical interest. The first total absorption detection systems to be
constructed
with low neutron sensitivity were large liquid scintillation tanks of a
volume ranging from 300 to 3000 liters viewed by photomultiplier tubes
and with a through--hole to place the capture sample. However, those systems
had serious drawbacks.
The absorption of all $\gamma$--radiation cannot be achieved. The peak around
the
excitation energy in the capture process has a large tail towards the lower
energies
due to partial escape of radiation. The 2.2\,MeV background from hydrogen
capture in the
scintillator limits the determination of the peak fraction below. Therefore, an
efficiency of typically 70\,\% for the detection of the capture events is
obtained.

The idea of total absorption detection has been further improved by using
a ball of scintillation crystals viewed by photomultiplier
tubes instead of a liquid scintillation tank. These crystal ball type
detectors were first constructed with NaI(Tl) crystals
to measure $\gamma$--ray cascades in heavy ion reactions.
Using BaF$_2$ crystals the
crystal ball detector was also suitable for neutron capture measurements.
The neutron sensitivity is considerably lower than that
of NaI(Tl), fast timing is possible because of a fast component in the
scintillation
light (600\,ps decay time), and the energy resolution is comparable to that
of NaI(Tl).
At the Karlsruhe 3.75\,MV Van de Graaff accelerator such a detector has
been constructed
and successfully used to determine neutron capture cross sections. The detector
consists of 42 BaF$_2$ crystals coupled to photomultiplier tubes, 12
pentagon and 30 hexagon crystals, which form a spherical shell
of 15\,cm thickness. The loss in 4$\pi$ solid angle due to the openings for the
neutron beam, the samples, and leaks between the crystals is less than 5\,\%.
The total efficiency in the measurements is close to 95\,\%.
The accuracy claimed in these
capture measurements is as good as 1\,\%. Details of this detector are well
documented~\cite{wis90} and a series of astrophysically important
isotopes has been
measured~\cite{wis92,wis93,vos94,wis95}. Elements were selected with two or three s--only
isotopes in the isotopic chain (e.g., $^{122,123,124,125,126}$Te, $^{134,135,136,137}$Ba,
$^{147,148,149,150,152}$Sm, $^{152,154,155,156,157,158}$Gd). However,
because the Van de Graaff accelerator as a white neutron
source for time--of--flight measurements cannot provide enough neutrons in the energy
range from a few eV to about 5\,keV the measurement of the excitation
function of an isotope must remain incomplete.
Consequently the Maxwellian--averaged capture (MAC)
cross sections cannot be determined at the temperature
of the $^{13}$C($\alpha$,n) astrophysical neutron source
($k_{\rm B}T=$8--12\,keV) without relying
on supplementing cross--section data from literature. In the s--process
the crucial isotopes are the so--called bottle\-neck isotopes with magic
neutron shells and well--resolved resonance structure up to 100\,keV.
For the measurement of these species another accelerator and detection
method, the total energy detection, is preferred. In these measurements the
required energy range is completely covered and the resonance strengths
are determined with an optimum signal--to--background ratio.

\subsubsection{Total Energy Detection}
Systems for total energy detection in use are the Moxon--Rae detector and a
generalization of this detection principle to any $\gamma$--ray detection system.
The principle of these detector types is opposite to the total absorption
detectors. The aim is not to sum up the $\gamma$--ray cascades of the capture
process but to detect not many more than one of the
emitted photons of the cascade from each capture event.
Mainly, this requires a low efficiency detector.
The Moxon--Rae detector consists of a graphite disc for converting $\gamma$--rays
into electrons followed by a thin plastic scintillator coupled to a phototube
to detect the electrons. In this way the efficiency ($\le$1\,\%)
becomes proportional
to the detected $\gamma$--ray energy $E_{\gamma i}$. Accumulating the capture
events with such a detector then results in an overall efficiency in the capture
measurement of
\begin{eqnarray}
\varepsilon=\sum_{i=1}^m \varepsilon_i(E_{\gamma i})=
k \sum_{i=1}^m E_{\gamma i}=k E^* \quad ,
\end{eqnarray}
which is proportional to the excitation energy of the capture process, and, therefore,
independent of the details of the individual cascades. This simple detector
principle has been extensively used for measurements because of its fast timing
abilities and the low sensitivity to scattered neutrons.
Unfortunately for Moxon--Rae detectors
the proportionality of the $\gamma$--efficiency to $\gamma$--energy is
only fulfilled approximately.

However,
the principle of total energy detection can be applied to each kind of
detector
system provided the detector response $R(I,E_{\gamma}$), i.e., the probability
that a $\gamma$--ray of energy $E_{\gamma}$ gives rise to a pulse of amplitude $I$,
is transformed on-- or off--line by a so--called weighting function W(I)
to an efficiency proportional to the detected photon energy $E_{\gamma}$:
\begin{eqnarray}
\int_{I_{\rm l}}^{I_{\rm u}} R(I,E_{\gamma})W(I)dI=E_{\gamma} \quad .
\end{eqnarray}
The integration limits $I_{\rm l}$ and $I_{\rm u}$ are chosen to cover the expected
pulse height amplitudes in the capture process.
This elegant generalization of the Moxon--Rae detector was first applied by
Macklin and Gibbons~\shortcite{mac67} to a detector system consisting of a pair of
C$_6$F$_6$ liquid scintillation detectors.
In this way the efficiency of the detector could be increased
to about 15\,\% but the use of this detector for capture measurements
depends on an accurate determination of the weighting function, chiefly a
property of the detector system. The new detector keeps the advantages of the
Moxon--Rae detector, good timing properties and low neutron sensitivity, and
improves it in efficiency and in the property of total energy detection.
Instead of the C$_6$F$_6$ scintillator, C$_6$D$_6$ with a lower
neutron sensitivity is used now.

One persisting problem with these detectors --- the accurate
determination of the weighting function~\cite{cov95} --- has been solved
for the detector system in use at the electron linear accelerator GELINA
at Geel. It turned out that the weighting function had been calculated
from Monte--Carlo simulations
with an in\-ad\-e\-quate treatment of the electron transport and the effect of
the structural material of the sample--detector configuration. When the
weighting function was based
entirely on experimentally determined response
functions and efficiencies~\cite{cor91} a serious
problem found in the measurement of the resonance parameters of the 1.15\,keV
$^{56}$Fe resonance was suddenly solved.
Eventually, the GELINA results~\cite{per88} were also confirmed on the
whole, using a weighting function determined from
improved up--to--date computer calculations.

To attack astrophysical problems, the GELINA capture detector system was used
in measurements on $^{138}$Ba~\cite{bec94} and $^{208}$Pb~\cite{cor95}.
Previous $^{138}$Ba capture measurements had been
performed in a limited energy interval. The time--of--flight measurement
done at ORELA had a lower energy bias of 3\,keV~\cite{mus79} and the activation
measurement~\cite{bee80} yielded a
MAC cross section only at $k_{\rm B}T$=25\,keV. The excitation function was
determined from a few eV to 300\,keV neutron energy.
Two strong p--wave resonances were
found below 3\,keV that affect the MAC cross section at $k_{\rm B}T$=8--12\,keV.
The calculated MAC cross section contains a 10\,\% contribution from direct
capture, estimated by theoretical calculations~\cite{bal94}.
\begin{table}
\caption[dub]{\label{t1}Resonance parameters and capture areas of
$^{208}$Pb+n in the range 1--400\,keV~\cite{cor95}.}
\begin{center}
\begin{tabular}{cccccc}
\hline
$E_0$ & $l$ & $J$ & $\Gamma_{\rm n}$& $\Gamma_{\gamma}$&
$g\Gamma_{\rm n}\Gamma_{\gamma}$/$\Gamma$\\
(keV) & & & (eV) & (meV) &  (meV)\\
\hline
43.34&--&--&--&--& 26.5$\pm$0.8 \\
47.33&--&--&--&--& 38.5$\pm$1.2 \\
71.21&1&3/2&101$\pm$5 &12.4$\pm$2.0&24.8$\pm$4.0\\
77.85&1&3/2&958$\pm$10&125$\pm$30  & 250$\pm$60 \\
86.58&1&1/2&75.4$\pm$3.0&15.2$\pm$6.0&15.2$\pm$6.0 \\
116.78&1&3/2&317$\pm$6& 27$\pm$10 &55$\pm$20 \\
130.25&2&5/2&9.7$\pm$0.9& 101.7$\pm$4.0 &302$\pm$12 \\
153.31&1&3/2&10.5$\pm$1.0& 26.7$\pm$4.5 &53.2$\pm$9.0 \\
169.48&1&3/2&21.9$\pm$1.6& 73.9$\pm$3.0 &147$\pm$6 \\
193.69&--&--&--&--&277$\pm$10 \\
350.43&--&--&--&--&319$\pm$34 \\
359.14&--&--&--&--&253$\pm$44 \\
\hline
\end{tabular}
\end{center}
\end{table}

The time--of--flight measurement on $^{208}$Pb was motivated by a serious discrepancy of a
factor two between
an activation measurement at $k_{\rm B}T$=30\,keV~\cite{raz88} and
the time--of--flight measurement reported from
ORELA~\cite{mac77}. The level density of this doubly magic nuclide is
especially low. In the
measurement from GELINA no new resonances were found (Table~\ref{t1}) but the
strength of the resonances was much lower than previously reported.
It turned out
that in this case direct capture is as significant as compound capture
and the activation measurement~\cite{raz88} provided the correct total
capture cross section at $k_{\rm B}T$=30\,keV (see Sect.~5.2).

\subsection{Activation Methods}
The development of a special neutron activation method~\cite{bee80,bee94}
to measure
capture cross sections at the Karlsruhe 3.75\,MV Van de Graaff accelerator
has lead to a large number of measurements. The method is simple and
the measurements can be repeated easily to reproduce the results. With a
high--resolution Ge--detector the technique is selective
and, therefore, very sensitive. The measurements can be carried out with
samples of natural composition in many cases. Because of its sensitivity
cross sections of only a few $\mu$barn (1\,barn = 10$^{-24}$\,cm$^2$)
can be measured. This is important
for the investigation of the direct--capture mechanism.

The neutrons are generated by the $^7$Li(p,n) and T(p,n) reactions.
For energy points at 25\,keV and 52\,keV one can take advantage of the special
properties of these reactions at the reaction threshold. For energy
points above 100\,keV thin targets (full half--width 15--20\,keV) are
applied.

For many light isotopes and the isotopes at magic neutron shells direct
capture is a significant, sometimes even the dominant capture--reaction
process (\citeauthor{gru95}, \citeyear{gru95};
\citeauthor{meis95a}, \citeyear{meis95a}, \citeyear{meis95b}, \citeyear{meis96};
\citeauthor{bee95}, \citeyear{bee95};
\citeauthor{kra96}, \citeyear{kra96}).
As direct capture yields a smooth cross section of less than 1\,mbarn in those
cases,
the common experimental time--of--flight techniques~\cite{bec94,cor95}
are normally not sensitive enough for a
measurement. This is different for the activation technique, especially
for the fast cyclic activation technique~\cite{bee94} which had to be
applied to perform the direct--capture measurements on the light isotopes.

\subsubsection{Common Activation Technique}
A normal activation measurement is subdivided into two parts:
(1) the irradiation of the
sample, (2) the counting of the induced activity~\cite{bee80}.
The characteristic time constants of an activation are the irradiation time
$t_{\rm b}$, the counting time $t_{\rm c}$ and the waiting time $t_{\rm w}$.
The activities of the samples were counted with a Ge--detector through the characteristic
$\gamma$--ray lines of the individual isotopes.
The $\gamma$--ray line
intensity $C_{\rm f}$ is given as
\begin{eqnarray}
\label{equ1}
C_{\rm f}=\epsilon_\gamma K_\gamma f_\gamma [1-\exp(-\lambda t_{\rm c})]
\exp(-\lambda t_{\rm w})
N \sigma f_{\rm b} \int_0^{t_{\rm b}} \Phi(t)dt \\
\quad\mbox{with}\quad
f_{\rm b}=\int_0^{t_{\rm b}} \Phi(t) \exp(-\lambda t)dt/\int_0^{t_b} \Phi(t)dt
\quad .
\nonumber
\end{eqnarray}
The following additional quantities have been defined: $\epsilon_\gamma$:
Ge--efficiency, $K_\gamma$:
$\gamma$--ray absorption, $f_\gamma$: $\gamma$--ray intensity per decay,
N: the number of target
nuclei, $\sigma$: the capture cross section, $\Phi$: the neutron flux.
The quantity $f_{\rm b}$ is calculated from the registered flux history of
a $^6$Li glass monitor.

Eq.~(\ref{equ1}) contains only the unknown quantities $\sigma$
and the time integrated neutron flux
$\int_0^{t_{\rm b}} \Phi(t)dt $. Therefore, cross section ratios can be formed
for different isotopes exposed to the same total neutron flux.
This is the basis for the determination
of the wanted capture cross section relative to the well--known standard
$^{197}$Au capture
cross section~\cite{rat88}. As the sample of the isotope to be investigated is
normally characterized by a non--negligible finite thickness
it is desirable to sandwich the sample between
two comparatively thin gold foils for
the determination
of the effective neutron flux at the sample position.
The activities of these gold foils
are counted individually.

In general, only a fraction of the activity $f_\gamma$ decays through the
selected line.
If this fraction is very small or if no $\gamma$--ray emission accompanies the
beta--decay it is necessary to detect the activity of the product
nucleus with
a 4$\pi$ beta--spectrometer. Important examples investigated are
$^{88}$Sr and $^{89}$Y~\cite{kae90} and $^{208}$Pb~\cite{raz88}.

The high sensitivity of the activation technique allows for measurements
with sample amounts of $\mu$g, provided the capture cross section of the
investigated isotope is of the order of one barn.
So far, the radioactive nuclei
$^{163}$Ho and $^{155}$Eu have
been studied~(\citeauthor{jag95a}, \citeyear{jag95a}, \citeyear{jag95b}),
which are important for s--process nucleosynthesis.
To carry out activations on radioactive isotopes the sample preparation
is the main problem. This application is promising and not yet fully exploited.

\subsubsection{Fast Cyclic Activation Technique}
The cyclic activation method is the repetition of the irradiation and activity
counting procedure of a normal activation for many times to gain statistics. Especially
for nuclei with half--lives of only minutes or seconds
a large number of irradiation and counting cycles is needed.
The technical details of this method can also be found in Appendix~\ref{A1}.

The cyclic activation technique is applicable to radioactivities of short
and long half--lives as well~\cite{bee94}.
The common activation technique is contained in the cyclic activation
technique as a special case which is
obtained if we choose only one correspondingly long
cycle. Eqs.~\ref{eq1} and \ref{eq2} both reduce to the formula for
the common activation technique (Eq.~\ref{equ1}).
The cyclic activation method is free
of saturation effects which limit the reasonable irradiation time of
a common activation
to about four times the half--life of the generated isotope.
With the cyclic activation
technique the optimum in
statistics can be obtained in all cases.

For many light isotopes direct
capture is the dominant capture reaction
process. This dominance has been found in cyclic measurements
of $^{14}$C~\cite{bew92},
$^{15}$N~\cite{meis96}, $^{18}$O~(\citeauthor{meis95a}, \citeyear{meis95a},
\citeyear{meis95b})
$^{22}$Ne~\cite{bee91}, and $^{36}$S~\cite{bee95}.

\section{Theoretical Models and Calculations}
\subsection{Introduction}
\begin{table}
\caption{\label{t2}Reaction mechanisms and models.}
\begin{center}
\begin{tabular}{|c|c|}
\hline
Direct reaction (DI) & Compound nucleus (CN)\\
\hline
No CN levels & Many CN levels \\
Short interaction time &
Long interaction time\\
(10$^{-21}$--10$^{-22}$\,s) & (10$^{-14}$--10$^{-20}$\,s)\\
One--step process & Many--step process\\
Single--particle resonances & CN resonances\\
DI models (DWBA, DC), & Phenomenological models\\
Microscopic methods &  (Breit--Wigner, R--matrix),\\
(RGM, GCM, few--body) & Hauser--Feshbach (HF)\\
\hline
\end{tabular}
\end{center}
\end{table}
Nuclear burning in explosive astrophysical environments produces
unstable nuclei which again can be targets for subsequent reactions. In
addition, it involves a
very large number of stable nuclei, which are not fully explored
by experiments. Thus, it is necessary to be able to predict reaction
cross sections and thermonuclear rates with the aid of theoretical models.

In astrophysically relevant nuclear reactions two important reaction
mechanisms take place. These two mechanisms are compound--nucleus reactions
(CN) and direct reactions (DI).
\begin{enumerate}
\item The CN mechanism was proposed
about 60 years ago by N.~Bohr \cite{Boh36,Kal37}.
In this mechanism the projectile
merges with the target nucleus and excites many degrees of freedom
of the compound nucleus. The de--excitation proceeds by a multistep process
and therefore has a reaction time typically of the order
of $10^{-14}$\,s to $10^{-20}$\,s. After this time the compound nucleus
decays into various exit channels. The relative importance
of the decay channels is determined by the branching ratios
to the final states.
\item The DI mechanism has been introduced by
\citeauthor{But50} (\citeyear{But50,But51}).
In this case
the reaction proceeds in a single step from the initial to the final
state and has a characteristic time scale of about $10^{-21}$\,s to $10^{-22}$\,s.
This corresponds to the time the projectile needs to pass through
the target nucleus.
\end{enumerate}

A characterization and classification of the main reaction mechanisms and models
that are appropriate for nuclear astrophysics is given in Table~\ref{t2}.
The reaction mechanism and therefore
also the reaction model depends on the number of levels in the CN.
If one is considering only a few CN resonances the R--matrix theory
is appropriate.
In the case of a single resonance the R--matrix theory reduces to the
simple phenomenological Breit--Wigner
formula. If the level density of the CN is so high that there are
many overlapping resonances,
the CN mechanism will dominate and the statistical Hauser--Feshbach
method can be applied. Finally, if there are no CN resonances
in a certain energy interval the DI mechanism dominates and
one can use DI models or microscopic methods. Only in a few cases
resonances have a dominant single--particle structure:
such resonances are then called single--particle
resonances. In this case the resonance is not of CN nature
and can also be described in
the DI mechanism. In the following subsections we will briefly discuss the
reaction models that are mainly used in nuclear astrophysics.

\subsection{Phenomenological methods}
The most important phenomenological methods used in analyzing nuclear
reactions are the R--matrix theory and the simpler Breit--Wigner formula
for CN resonances.

In the R--matrix
formalism the cross section $\sigma_{if}(j,o;E)$
for the reaction $i(j,o)f$ describing CN resonances $\lambda$ with
angular momentum quantum number $J_\lambda$, parity $\pi_\lambda$, excitation energy
$E_\lambda$, partial widths of the entrance channel $\Gamma_{i \lambda}$
and exit channel
$\Gamma_{f \lambda}$, and the total width
$\Gamma_\lambda = \sum_c\Gamma_{c,\lambda}$ being the sum
over all channels $c$ is given by~\cite{Lane58}:
\begin{equation}
\label{RM}
\sigma_{if}(j,o;E)=
\frac{\pi \hbar^2}{2 \mu_{ij} E}\frac{1}{(2J_i+1)(2J_j+1)}
\sum_{\lambda} (2J_\lambda+1)
\frac {\Gamma_{i,\lambda} \Gamma_{f,\lambda}}
{(E_\lambda + \Delta_\lambda - E)^2 + \frac{\Gamma_\lambda^2}{4}} \quad ,
\end{equation}
where $E$ is the
center of mass energy and  $\mu _{ij}$ is the reduced mass. The energy shift
$\Delta_\lambda$ is the so--called Thomas--Lane correction and
represents the difference between the energy at which
the resonance is observed and the corresponding state
in the CN system. The Thomas--Lane correction results from the background terms of
the other resonances that are superimposed
on the considered resonance $\lambda$.

Even though the mathematical formalism of the R--matrix theory is perfectly
general, it is particular suited for CN reactions, because the resonances can
be identified with CN states. Direct reactions on the other hand can
be described in the language of the R--matrix theory only as a correlation
between many such states.
The R--matrix theory has been employed often in nuclear astrophysics, especially
in analyzing resonance structures of cross sections in the thermonuclear
energy region.

\subsection{Breit--Wigner formula}
In the case of a single resonance Eq.~\ref{RM} reduces to the well--known Breit--Wigner
formula~\cite{Bre36,Bla62}:
\begin{equation}
\label{BW}
\sigma_{if}(j,o;E) =
\frac{\pi \hbar^2}{2 \mu_{ij} E}
\frac{\left(2J+1\right)}{\left(2J_i+1\right)\left(2J_j+1\right)}
\frac{\Gamma_i \Gamma_f}
{\left(E_\lambda - E\right)^2 + \frac{\Gamma^2}{4}} \quad ,
\end{equation}
where $J$ is the angular momentum quantum number and $E_\lambda$ is the
excitation energy of the resonant state.
The partial widths of the entrance and exit channels
are $\Gamma_i$  and $\Gamma_f$, respectively.
The total width $\Gamma= \sum_{c^\prime}\Gamma_{c^\prime}$
is the sum of the partial widths over all channels $c^\prime$.

The Breit--Wigner formula has been employed extensively in nuclear astrophysics
for analyzing single isolated resonances. One important aspect is that
the partial width $\Gamma_c$ can be related to
spectroscopic factors $S_c^p$ for a particle $p$ in a state $c$ by
\begin{equation}
\label{SF}
\Gamma_c = C^2 S_c^p \Gamma_c^p \quad,
\end{equation}
where $C$ is the isospin Clebsch--Gordan coefficient.
The single--particle width $\Gamma_c^p$ can be calculated
from the scattering phase shifts of a scattering potential with
the potential depth determined by matching the resonance energy.

The partial widths can be calculated  with the help of Eq.~\ref{SF}
by using spectroscopic factors obtained
in other reactions, e.g., the spectroscopic factors
necessary for calculating the neutron partial widths
in A(n,$\gamma$)B can be extracted
from the reaction A(d,p)B. Also $\gamma$--widths can be
extracted indirectly from reduced electromagnetic transition
probabilities, e.g., the gamma widths in A(n,$\gamma$)B
can be obtained from electromagnetic transitions
of the nucleus B. For unstable nuclei
where only limited or even no
experimental information is available, the
spectroscopic factors and electromagnetically reduced transition probabilities
can also be extracted from nuclear structure theories (e.g., shell model).

\subsection{The Statistical Model (Hauser--Feshbach)}
In general, intermediate mass
and heavy nuclei have intrinsically
a high density of excited states due to their large nucleon number.
A high level density in the
CN at the appropriate excitation energy allows to
make use of the statistical model approach for compound nuclear
reactions (e.g., \citeauthor{Hauser52},~\citeyear{Hauser52};
\citeauthor{Mahaux79},~\citeyear{Mahaux79};
\citeauthor{Gadioli92},~\citeyear{Gadioli92}),
which averages over resonances. The statistical model approach has been
employed in calculations of thermonuclear reaction rates for astrophysical
purposes by many researchers, starting
with \citeauthor{Truran66}~\shortcite{Truran66},
\citeauthor{Michaud70}~\shortcite{Michaud70} and
\citeauthor{Truran72}~\shortcite{Truran72}, who only
made use of ground state properties.
\citeauthor{Arnould73}~\shortcite{Arnould73} pointed out the
importance of excited states of the target. Presently, the compilations
by \citeauthor{Holmes76}~\shortcite{Holmes76},
\citeauthor{Woos78}~\shortcite{Woos78} and
\citeauthor{Thiele87}~\shortcite{Thiele87}
are the ones utilized
in large scale applications in all subfields of nuclear astrophysics
when experimental information is unavailable.

A (sufficiently) high level density in the compound nucleus permits
to use averaged
transmission coefficients $T$, which do not reflect a resonance
behavior,
but rather describe absorption through an imaginary part in the (optical)
nucleon--nucleus potential (for details see
\citeauthor{Mahaux79},~\citeyear{Mahaux79}).
This leads to the well--known Hauser--Fesbach expression given in
Appendix~\ref{A2}.

\begin{figure}
\centerline{\psfig{figure=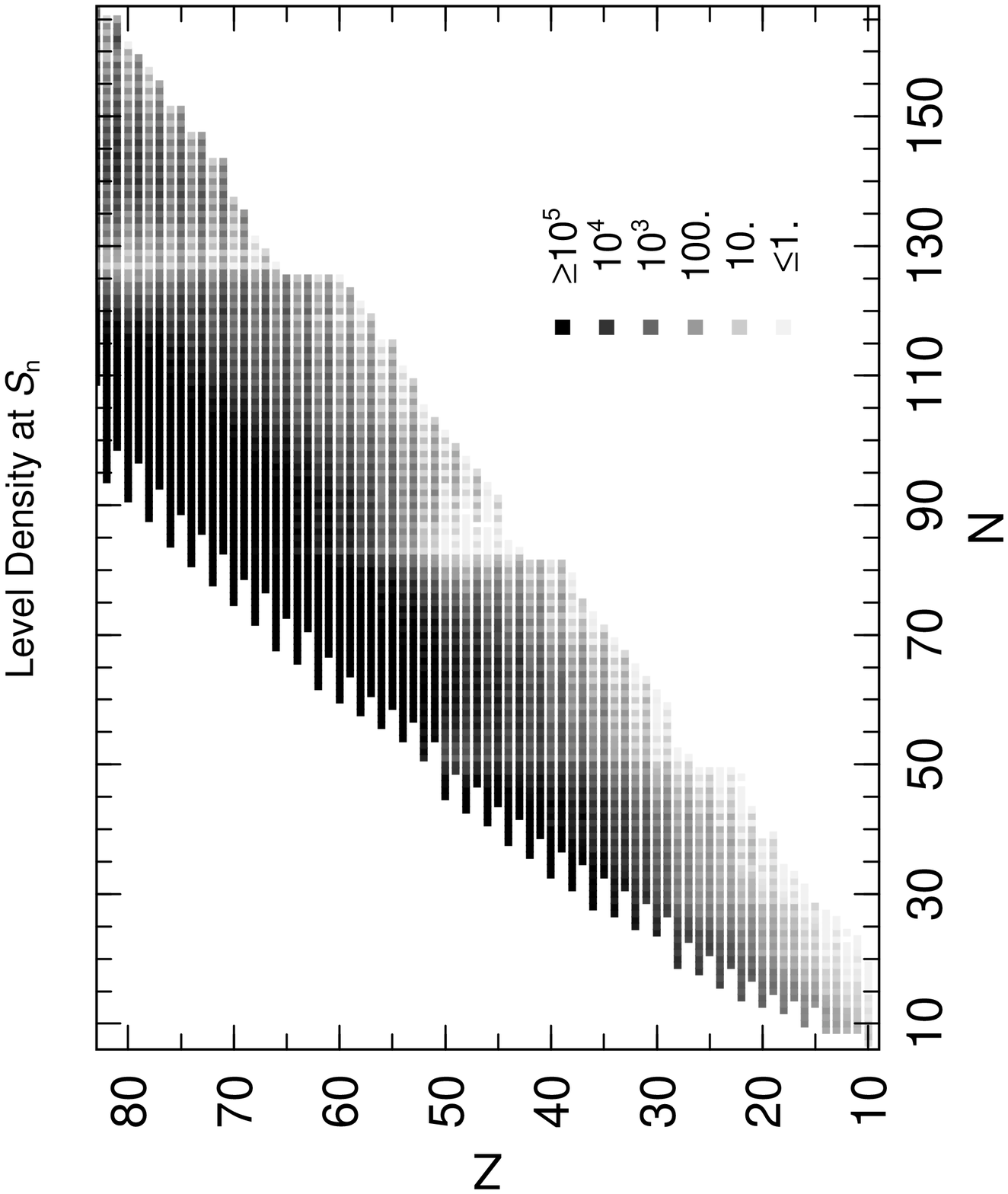,width=13cm}}
\caption{\label{f1} Level density (in levels per MeV) at the
respective neutron separation
energy \protect\cite{Rau95b,Rau96}.}
\end{figure}

Considerable effort has been put into providing the best possible input
functions for the statistical model calculations. Entering Eq.~\ref{hf}
are particle-- and photon--transmission coefficients, width fluctuation
corrections, masses, and level densities.
The level densities were subject to the largest part of the
uncertainties in the  most recent cross--section calculations~\cite{Cowan91}.
It is hoped that the accuracy of the theoretical predictions will be
further improved by employing advanced level density descriptions
(see Sect.~5.3).

It is often colloquially termed that the statistical model is
only applicable for intermediate and heavy nuclei. However, the only
necessary
condition for its application is a large number of resonances
at the appropriate bombarding energies, so that the cross section can be
described by an average over resonances. This can be
valid for light nuclei
in specific cases and on the other hand not valid for intermediate mass
nuclei near magic numbers.
In the case of neutron--induced reactions a criterion for the
applicability can directly be derived from the level density. For
astrophysical purposes the projectile energies are quite low by nuclear
physics standards. Therefore, the relevant energies will lie very close
to the neutron separation energy. Thus, one only has to consider the
level density at this energy. As a rule--of--thumb it is usually said that
there should be at least 10 levels per MeV for reliable statistical
model calculations. The level densities at the appropriate neutron
separation energies are shown in Fig.~\ref{f1} (note that therefore
the level density is plotted at a {\em different} energy for each
nucleus). One can easily identify the magic neutron numbers by the drop
in level density, as well as odd--even staggering effects. A general
sharp drop is also found for nuclei close to the neutron drip line. For
nuclei with such low level densities the statistical model cross
sections will become very small and other processes might become
important, such as direct reactions.

The above plot can give hints on when it is safe to use the statistical
model approach and which nuclei have to be treated with special
attention at a given temperature. Thus, information which nuclei
might be of special interest for an experimental investigation may also
be extracted.
However, such plots can only give
clues as to which reactions have to be studied more carefully as, e.g.,
the ``sufficient number of levels'' is only an estimate.
Reactions in the temperature range $0 \le T_9 \le 0.01$\,should always be
treated with special care because of the possible importance of single
resonances (which can be included in the CN calculation
when known; cf., Eq.~\ref{trans}).

\subsection{Microscopic methods}
Microscopic methods like the Resonating Group Method (RGM) or
Generator Coordinate
method (GCM) start from nucleon--nucleon interactions and are based on
many--nucleon
wave functions of the nuclei involved. In this approach, the explicit inclusion
of the Pauli principle leads to highly nonlocal potentials for the interaction
between the composite nuclei in the entrance and exit channel. It is obvious
that a fully microscopic approach like RGM and GCM is more satisfying,
since it is a first--principle approach.

The main drawback of the RGM is that it requires extensive analytical
calculations without systematic character when going from one reaction
to another. Consequently, the application of the RGM is
essentially restricted to reactions involving only a small number
of nucleons. This problem can be overcome by the GCM
\cite{Hil53}. In the GCM the relative wave functions
are expanded in a Gaussian basis. The GCM is similar to the RGM,
but it allows systematic
calculations, well suited for a numerical approach.

A microscopic model
starts from the nucleon--nucleon interactions and should not contain any free
parameter. Therefore, it is possible to predict physical properties of
the system
independently of experimental data. However, in most cases the fully microscopic
approach rarely reproduces physical observables to which
the calculations of astrophysical cross sections are sensitive, such
as thresholds,
resonance and bound--state energies or scattering data. Even if
the nucleon--nucleon interaction in microscopic calculations
is fitted to reproduce one such relevant observable, other
sensitive observables
often cannot be reproduced simultaneously. Nevertheless, microscopic
methods have also been used extensively for the investigation of
astrophysical relevant cross sections and many interesting
results have been obtained.  Review articles on
microscopic theories including examples of astrophysical processes
are found in
\citeauthor{Lan86} (\citeyear{Lan86}, \citeyear{Lan88}, \citeyear{Lan91}),
\citeauthor{Bay89}
\shortcite{Bay89}, \citeauthor{Obe91} \shortcite{Obe91},
\citeauthor{Des93} \shortcite{Des93}, and \citeauthor{Lan96}
\shortcite{Lan96}.

\subsection{Direct--reaction models}
Direct--reaction (DI) models are based on the description of the
dynamics of the reaction by a Schr\"odinger equation with local optical
potentials in the entrance and/or exit channels.
Such models are the {\em Distorted Wave Born Approximation} (DWBA)
\cite{Aus70,Sat83,Gle83,Obe91}
for transfer or the {\em Direct Capture} model (DC)
\cite{Chr61,Tom63,Rol73,Obe91} for capture reactions.
The expressions for the DWBA and DC cross sections
are given in Appendix~\ref{A3}.

The DI models can be derived from microscopic theories
essentially by allow\-ing approximations in the antisymmetrization
procedure \cite{Wil77}. In principle, DI models neglect
the antisymmetrization of the optical potentials in the entrance and
exit channel as well as exchange processes. However, the
effects of the Pauli principle can be taken into account
phenomenologically by fitting the parameters of
the optical potentials to elastic scattering data or to phase shifts
obtained from microscopic models. Furthermore, also exchange
processes such as knock out and heavy--particle pickup and stripping have
to be considered in some cases. This may be necessary, e.g., to describe
differential reaction cross sections on light target nuclei.

The DWBA and DC are based on the premise that elastic scattering in the
entrance and exit channel is dominant compared to the flux into
other channels. The DWBA and DC are well established
for higher projectile energies ($\ge10-20$\,MeV) and transitions
to low--lying states of the residual nucleus. For
lower energies precompound (e.g., exciton) or compound (e.g.,
Hauser--Feshbach) models have to be used. However, as stated before, for
thermonuclear or thermal energies target nuclei
can sometimes have very low level densities.
In these cases the contributions from the CN mechanism
can be small and the DI mechanism
cannot be neglected and may even dominate the reaction.

Important for the success of the DI models is the fact that
the optical potentials are taken from
realistic
models, i.e., from semi--microscopic formalisms such as the
folding--potential model,
and not only from empirical potentials like Saxon--Woods potentials.
Folding potentials have been used extensively and successfully in
potential--model calculations for astrophysically relevant
nuclear reactions.
The folding procedure is used for calculating the real parts of the
optical or bound--state potentials to describe the
elastic scattering data or the bound states, respectively. In most
astrophysical applications one of the particles is a nucleon (proton or
neutron), $\alpha$--particle, deuteron, triton or helion.
In the folding approach the nuclear density
$\rho_{A}$ is folded with an energy and density dependent nucleon--nucleon
interaction $v_{\rm eff}$ \cite{Kob84,Obe91}. For a
nucleon--nucleus system we use single folding
\begin{equation}
\label{SFo}
V(R) = \lambda \int \; d{\bf r}
\rho_{A} ({\bf r}) v_{\rm eff}(E,\rho_A,|{\bf R}  - {\bf r}|)
\end{equation}
and, for a system with both interacting nuclei having mass numbers $A \ge 2$,
double folding
\begin{equation}
\label{DFo}
V(R) = \lambda \int \; d{\bf r_1} d{\bf r_2}
\rho_{A} \left({\bf r_1}\right) \rho_{a} \left({\bf r_2}\right)
v_{\rm eff}\left(E,\rho_a,\rho_A,\left|{\bf R} -
{\bf r_1} - {\bf r_2} \right|\right) \quad .
\end{equation}
In the above expressions $A$ and $a$ are the two colliding nuclei and
${\bf {R}}$ is the separation of their centers--of--mass.
The normalization factor $\lambda$ is adjusted to reproduce the
experimental elastic--scattering and separation--energy data
for the different channels involved in the considered reaction.
This is of special importance for the calculation
of astrophysically relevant cross sections, because
the correct results of such calculations depend
sensitively on the reproduction of the above observables.
Also,
because at low energies only a few channels are open (sometimes
only two or three), only a small or even no
imaginary part of the optical potentials, describing absorption
into other channels, is necessary.

\section{Selected examples}
In this section we consider three selected examples:
neutron capture on $^{36}$S~\cite{bee95,Ober95},
$^{208}$Pb~\cite{cor95} and on Gd--isotopes~\cite{wis95,Rau95a}.

The main emphasis will be given to the
first reaction, because in this case the reaction mechanism is
predominantly direct and is therefore sensitive to the
details of nuclear structure. Often such reactions
have been investigated erroneously using the Hauser--Feshbach method.
However, as already stated before, if there are no or only
a few CN levels in the energy region of interest,
the direct reaction mechanism can
even dominate the reaction. There are other recent examples
for neutron capture that have been investigated
and show a dominance of the direct mechanism:
e.g., $^{12}$C(n,$\gamma$)$^{13}$C~\cite{ots94,men95},
$^{15}$N(n,$\gamma$)$^{16}$N~\cite{meis96},
$^{18}$O(n,$\gamma$)$^{19}$O~\cite{gru95,meis95b}, and
$^{48}$Ca(n,$\gamma$)$^{49}$Ca~\cite{kra96,bee96}.
Since the
reaction $^{36}$S(n,$\gamma$)$^{37}$S involves nuclei
at the border of the region of stability, it is also
a benchmark for nuclear--structure calculations necessary
for nuclei far--off stability.

In the reaction $^{208}$Pb(n,$\gamma$)$^{209}$Pb
the DI and CN mechanism give similar values
for the cross section in the
astrophysical energy range. We compare the resonant
and non--resonant contributions of the capture
cross section at thermonuclear energies with
experimental data. There are other examples for
neutron capture by magic target nuclei (e.g., $^{86}$Kr,
$^{88}$Sr, $^{136}$Ba, $^{138}$Ba), where direct
capture cannot be neglected \cite{bal94}.

The usual method to calculate neutron--induced
reactions at thermonuclear energies is the statistical Hauser--Feshbach method
(cf., Sect.~4.4).
However, as stated before, this is only correct
if the level density is high enough and the CN mechanism
is dominating. This will be the case for the majority of nuclear reactions
to be considered in nuclear astrophysics.
As one example for a dominating CN mechanism we
discuss neutron capture by Gd--isotopes at thermonuclear energies and
compare the values for the cross section obtained
in the Hauser--Feshbach formalism
with the experimental values.
\begin{table}
\caption[36S]{\label{t3}Final states, Q--values, transitions and cross
sections for $^{36}$S(n,$\gamma$)$^{37}$S
at 25.3\,meV, 25\,keV, 151\,keV, 176\,keV, and 218\,keV using DC
with the experimental data.}
\begin{center}
\begin{tabular}{cccrrrrr}\hline
Final & Q--value & Transition &
\multicolumn{5}{c}{Cross section} \\
state & [MeV] & &
25.3\,meV & 25\,keV & 151\,keV & 176\,keV & 218\,keV \\
&&&
[mbarn] & [$\mu$barn] & [$\mu$barn] & [$\mu$barn] & [$\mu$barn] \\
\hline
$\frac{7}{2}^-$ & 4.303 & d $\rightarrow$ f &
0.0 & 0.0 & 0.1 & 0.2 & 0.2 \\
$\frac{3}{2}^-$ & 3.657 & s $\rightarrow$ p &
157.1 & 158.0 & 64.3 & 59.5 & 53.5 \\
$\frac{3}{2}^+$ & 2.906 & p $\rightarrow$ d &
0.0 & 0.0 & 0.0 & 0.0 & 0.0 \\
$\frac{3}{2}^-$ & 2.312 & s $\rightarrow$ p &
5.3 & 5.3 & 2.1 & 2.0 & 1.8 \\
$\frac{1}{2}^-$ & 1.666 & s $\rightarrow$ p &
28.3 & 28.5 & 11.6 & 10.8 & 9.7 \\
\cline{4-8}
\multicolumn{3}{r}{Total cross section: DC} &
190.7 & 191.8 & 78.1 & 72.5 & 65.2 \\
\cline{4-8}
\multicolumn{3}{r}{Total cross section: experiment} &
$150 \pm 30$$^{1}$
&  $187 \pm 14$ & $81 \pm 7$ & $125 \pm 11$ & $78 \pm 7$ \\
\hline
\end{tabular}
\end{center}
{\footnotesize $^1$ \citeauthor{Sear92} \shortcite{Sear92}}
\end{table}

\subsection{Direct neutron capture by $^{36}$S}
%
%
%
\begin{figure}
\centerline{\psfig{figure=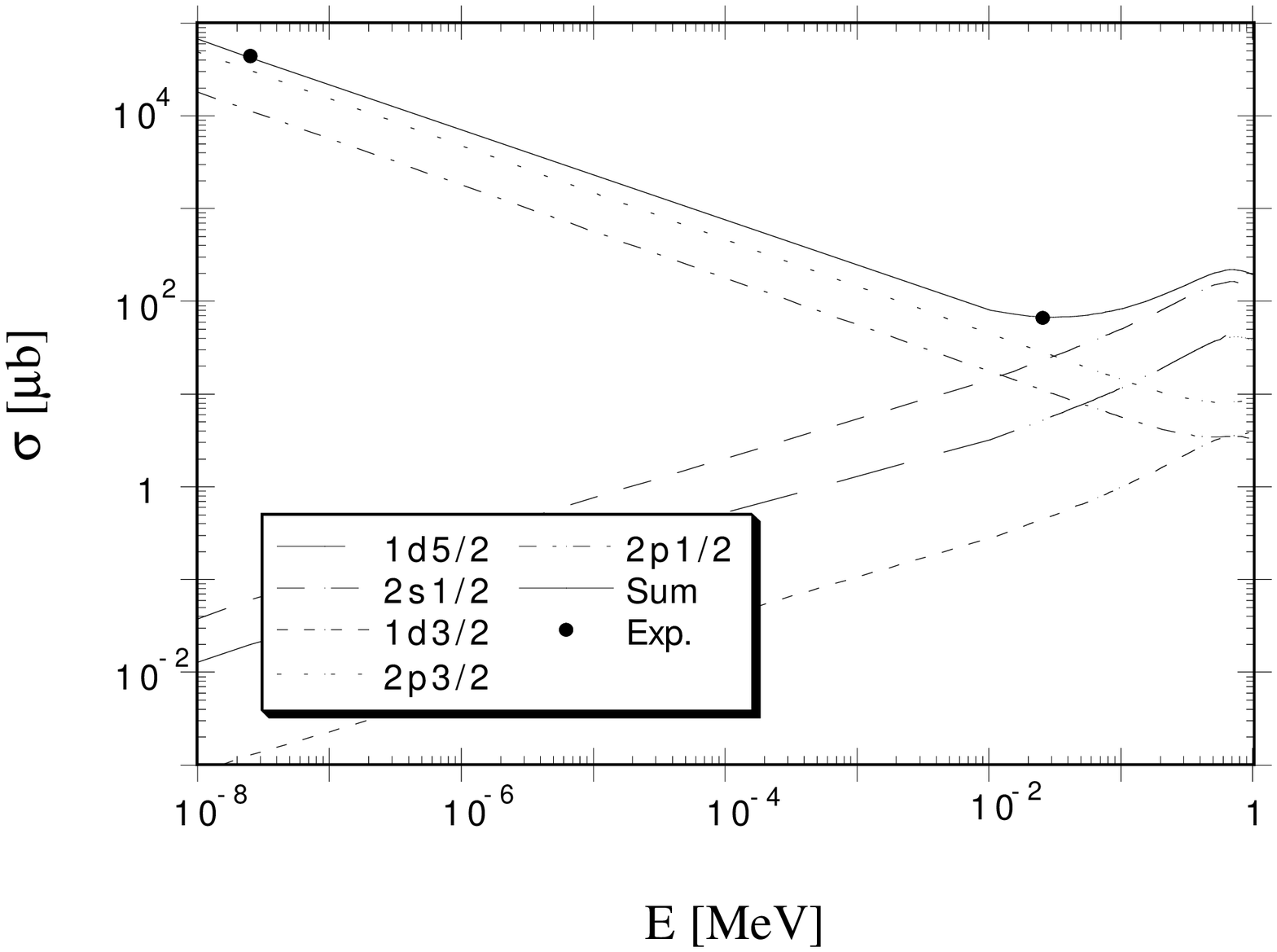,width=13cm}}
\caption[DC cross section]{\label{f2} Comparison of the DC cross section for
$^{36}$S(n,$\gamma$)$^{37}$S
with the experimental data from thermal to thermonuclear projectile energies.
The DC contributions for the different transitions to the final states of
$^{37}$S as well as the sum of all transitions (solid curve) are shown. The
experimental data at the thermal energy have been taken from
\protect\citeauthor{Sear92} \protect\citeyear{Sear92}.
}
\end{figure}
%
%
%

For the rare isotope $^{36}$S a significant abundance contribution is
expected from s--process nucleosynthesis. For quantitative
analyses the size of the destruction rate, the neutron capture rate,
is of fundamental importance
to estimate the magnitude of the $^{36}$S abundance formed by the weak and
main s--process components.
Using the statistical model the $^{36}$S capture cross
section has been estimated to be 300\,$\mu$barn at 30\,keV by
\citeauthor{Woos78}~\shortcite{Woos78} and 297\,$\mu$barn at 30\,keV by
\citeauthor{Cowan91}~\shortcite{Cowan91} (1\,eV = $1.60219\times 10^{-19}$\,J;
1\,barn = 10$^{-24}$\,cm$^2$).

We investigate the capture reaction $^{36}$S(n,$\gamma$)$^{37}$S
from thermal (25.3\,meV) to thermonuclear (25\,--218\,keV)
projectile energies
and compare the calculated cross sections in the DC--model
with the experimental data. The experimental
investigation was performed using
the fast cyclic activation technique~\cite{bee94} developed at the Karlsruhe
3.75\,MV Van de Graaff accelerator.

In the folding approach the nuclear density $\rho_{A}$ for the stable
nucleus
$^{36}$S was derived from the experimental charge
distribution~\cite{deVr87}.
The normalization factor $\lambda$ of the optical potential
in the entrance
channel was adjusted to fit the thermal
($^{36}$S+n)--scattering cross section of
($1.1 \pm 0.8$)\,barn~\cite{Sear92}.
Although this cross section is not determined well, we fitted
our normalization factor to reproduce 1.1\,barn.
However, applying the same fitting procedure to
the ($^{34}$S+n)--scattering cross section that is known
much better (($1.52 \pm 0.03$)\,barn,~\citeauthor{Sear92}, \citeyear{Sear92})
we obtained almost the same volume integral per nucleon in the two cases
($^{34}$S+n: 501.7\,MeV\,fm$^3$; $^{36}$S+n: 497.1\,MeV\,fm$^3$).
The
imaginary part of the optical potential is small for
the ($^{36}$S+n)--channel and can be neglected.
For the exit channels the normalization constants $\lambda$
were adjusted to the energies of the
ground and the excited states. The potentials obtained in
this way ensure the correct behavior of the wave functions in the nuclear
exterior.

The spectroscopic factors for one--nucleon stripping
of $^{37}$S were determined from the most recent
experimental $^{36}$S(d,p)$^{37}$S--data~\cite{Endt90}.
The masses and Q--values for the transitions
to the different states of the residual nucleus
$^{37}$S were taken from experimental data~\cite{Audi93,Endt90}.
For the DC--calculations the code TEDCA~\cite{TEDCA} was used.

The cross section for the reaction $^{36}$S(n,$\gamma$)$^{37}$S
obtained from the DC--calculation is compared to the
experimental data from the thermal to the thermonuclear
energy region in Fig.~\ref{f2}.
There are two types of E1--transitions contributing to
the transitions to the residual nucleus $^{37}$S.
The first one arises from an s--wave in the
entrance channel exciting the negative--parity states
3/2$^{-}$ and 1/2$^-$ (see Table~\ref{t3}).
These transitions give the well--known
1/v--behavior (see Fig.~\ref{f2}). The second type of E1--transition
comes from
an initial p--wave and excites the positive--parity state 3/2$^{+}$
in the final nucleus. This transition
has a v--behavior and can be neglected in the relevant energy range
(see Fig.~\ref{f2} and Table~\ref{t3}). The E1--transition to the
7/2$^-$--ground state
of $^{37}$S can be neglected also, because of the higher centrifugal
barrier
of the incoming d--wave (see Table~\ref{t3}). As can be seen
from Fig.~\ref{f2} this contribution affects the deviation from an
1/v--behavior of the cross section only above about 700\,keV.

The spin and parity assignments of the final states in $^{37}$S and
the Q--values for the transitions to the different final states
are shown in Table~\ref{t3}. Also the calculated
cross sections for $^{36}$S(n,$\gamma$)$^{37}$S
at 25.3\,meV, 25, 151,
176, and 218\,keV using DC with the
spectroscopic factors obtained from the (d,p)--data
are compared with the experimental data in this table.
Direct--capture calculations using the folding procedure
can  excellently reproduce the non--resonant experimental data
for the capture cross section by the neutron--rich sulfur isotope
$^{36}$S in the thermal and thermonuclear
energy region.
The enhancement in the region of 176\,keV
comes from resonant contributions~\cite{Endt90} not considered in the
DC--calculation.

We have determined the thermonuclear reaction rate
$N_{A}\left<\sigma v \right>$~\cite{Fowl67}. Since the cross
section follows an 1/v--law up to 150\,keV we obtain a
constant reaction rate (c.f., App.~A)
\begin{equation}\label{3}
N_{A}\left<\sigma v \right> = 2.56 \times
10^{4}\,{\rm cm}^3\,{\rm mole}^{-1}\,{\rm s}^{-1} \quad .
\end{equation}

The s--process production of $^{36}$S was recently discussed quantitatively
by \citeauthor{Scha95}, \shortcite{Scha95}
but without a reliable $^{36}$S(n,$\gamma$) cross
section.
The s--process reaction network in the sulfur to
calcium region contains (n,$\gamma$), (n,p), and (n,$\alpha$) reactions.
The $^{36}$S production is mediated by the $^{36}$Cl(n,p)$^{36}$S reaction
from seed nuclei with mass numbers $A<36$. But also seed nuclei $A>36$ can
contribute through the $^{39}$Ar(n,$\alpha$)$^{36}$S reaction channel.
Besides
its formation, the destruction of $^{36}$S by the
(n,$\gamma$) reaction is important. A decrease in the
$^{36}$S(n,$\gamma$)$^{37}$S
cross section leads to a corresponding increase of the abundance formed.
As our
measured $^{36}$S(n,$\gamma$)$^{37}$S  value is smaller by a factor of 1.8
than the
estimate of Woosley et al.~\cite{Woos78} the s--process abundance production
of $^{36}$S will be enhanced by this factor.
The quantitative analysis requires also model parameters for the main and
especially
the weak s--process component. This information can be obtained from the
analysis of the s--process beyond $A=56$~\cite{Beer95}.

DC could also be the
dominant reaction mechanism for neutron capture by neutron--rich isotopes
far--off stability occurring in the $\alpha$--rich freeze--out and r--process.
Leaving the line of stability, the Q--value and therefore the excitation energy
of the compound nucleus becomes lower,
leading to a substantial diminuition of the level density of
the compound nucleus. Thus,
the DC--contribution can become the dominating reaction mechanism.
Nuclear--structure models are indispensable for extrapolating
reaction rates to nuclei near and far--off the region
of stability, be\-cause only a limited or no
experimental information is available in this region.

The DC cross section of the reaction $^{36}$S(n,$\gamma$)$^{37}$S
can be considered
as a benchmark for different nuclear--structure models
(Shell Model, Relativistic
Mean Field Theory,
Hartree Fock Bogoliubov Theory) for calculating
neutron--capture cross sections by neutron--rich nuclei off--stability
taking place in the $\alpha$--rich freeze--out or r--process~\cite{Ober95}.
We find discrepancies of about a factor of two for the DC cross
sections with input parameters from the nuclear--structure models
in the case of $^{36}$S(n,$\gamma$)$^{37}$S. For astrophysical
applications in nuclear--network calculations such discrepancies
can be acceptable sometimes.
Clearly, investigations
of nuclear--structure models in the manner shown above are necessary to
test the reliability of such models for the application to calculations
of cross sections of astrophysically relevant nuclear reactions
involving nuclei far--off stability.

\begin{table}
\caption[dub]{\label{t6}The resonance and direct capture component of the
Maxwellian--averaged cross section $\sigma$ as a function of stellar temperature
$k_{\rm B}T$~\cite{cor95}.}
\begin{center}
\begin{tabular}{cccc}
\hline
$k_{\rm B}T$&\multicolumn{3}{c}{$\sigma$}\\
(keV)&\multicolumn{3}{c}{(mbarn)}\\
\cline{2-4}
&resonant& direct& total\\
&capture&capture&capture\\
&(exp)&(theory)&(exp+theory)\\
\hline
5 & 0.0015$\pm$0.0001&0.056$\pm$0.011&0.058$\pm$0.011 \\
8 & 0.018$\pm$0.001&0.070$\pm$0.014&0.088$\pm$0.014 \\
10 & 0.039$\pm$0.003&0.079$\pm$0.016&0.118$\pm$0.016 \\
12 & 0.063$\pm$0.005&0.086$\pm$0.017&0.149$\pm$0.018 \\
15 & 0.102$\pm$0.009&0.096$\pm$0.019&0.198$\pm$0.021 \\
17 & 0.126$\pm$0.012&0.102$\pm$0.020&0.228$\pm$0.023 \\
20 & 0.157$\pm$0.017&0.111$\pm$0.022&0.268$\pm$0.028 \\
25 & 0.196$\pm$0.023&0.124$\pm$0.025&0.320$\pm$0.034 \\
30 & 0.221$\pm$0.027&0.135$\pm$0.027&0.356$\pm$0.038 \\
35 & 0.235$\pm$0.029&0.145$\pm$0.029&0.380$\pm$0.041 \\
40 & 0.241$\pm$0.029&0.153$\pm$0.031&0.394$\pm$0.042 \\
45 & 0.243$\pm$0.025&0.162$\pm$0.032&0.405$\pm$0.041 \\
50 & 0.241$\pm$0.028&0.169$\pm$0.034&0.410$\pm$0.044 \\
60 & 0.231$\pm$0.026&0.180$\pm$0.036&0.411$\pm$0.044 \\
70 & 0.216$\pm$0.023&0.189$\pm$0.038&0.405$\pm$0.044 \\
80 & 0.201$\pm$0.021&0.194$\pm$0.039&0.395$\pm$0.044 \\
90 & 0.186$\pm$0.015&0.195$\pm$0.039&0.381$\pm$0.042 \\
100 & 0.171$\pm$0.017&0.195$\pm$0.039&0.366$\pm$0.043 \\
\hline
\end{tabular}
\end{center}
\end{table}
\begin{table}
\caption[Gd]{\label{crosstab}Experimental data and Hauser--Feshbach
calculations for Max\-well\-ian--averaged cross sections
at 30\,keV in mbarn for neutron capture on Gd--isotopes.}
\begin{center}
\begin{tabular}{lcc}
\hline
Target & \citeauthor{wis95}~\shortcite{wis95} (exp.) & Hauser--Feshbach (theor.) \\
\hline
$^{152}$Gd&1049$\pm$17 &870\\
$^{154}$Gd&1028$\pm$12 &622 \\
$^{155}$Gd&2648$\pm$30 &2340 \\
$^{156}$Gd&615$\pm$5 &455 \\
$^{157}$Gd&1369$\pm$15 &1426 \\
$^{158}$Gd&324$\pm$3&265\\
\hline
\end{tabular}
\end{center}
\end{table}
\subsection{Neutron capture by $^{208}$Pb}
The s--process of stellar nucleosynthesis terminates
at the isotopes of Pb and Bi, since all further neutron
capture leads to $\alpha$--unstable nuclei that are
then cycled back to the main lead isotopes.
To explain the abundances of these isotopes, the so--called
strong s--process component was introduced. To study its characteristics
in particular the neutron--capture cross section
of $^{208}$Pb is needed.

Recently, it became possible to determine the resonant part of the
capture cross section for
$^{208}$Pb(n,$\gamma$)$^{209}$Pb from experiment~\cite{cor95}. In that work high
resolution neutron capture measurements were carried out to
determine twelve resonances in the range 1--400\,keV~(Table
\ref{t1}; see also Sec.\ 3.1). From the data of Table~\ref{t1} the
resonant Maxwellian--averaged capture (MAC) cross sections (compound capture)
were calculated (second column of Table~\ref{t6}), using the Breit--Wigner formalism.

Using the experimentally known density distributions~\cite{deVr87},
masses~\cite{Audi93} and energy levels~\cite{Mar91}, we
calculated the non--resonant contribution in the DC model. The strength
parameter $\lambda$ of the folding potential in the entrance channel was
fitted to experimental scattering data at low energies~\cite{Sear92}.
The value of $\lambda$ for the bound state in the exit channel
is fixed by the requirement
of correct reproduction of the binding energies.
The spectroscopic factors for the relevant low lying states of $^{209}$Pb
are close to unity as can be inferred from
different $^{208}$Pb(d,p)$^{209}$Pb reaction data~\cite{Mar91}.
The results for the non--resonant MAC (direct capture) cross section
can then also be calculated (third column of Table~\ref{t6}).

The total MAC cross section is computed as the sum
of the resonant part determined from experiment and the non--resonant part
from theory (last column
of Table~\ref{t6}). The obtained total MAC cross section at 30\,keV
of ($0.356 \pm 0.038$)\,mbarn is in
excellent agreement with the value obtained from the activation experiment
($0.36 \pm 0.03$)\,mbarn of \citeauthor{raz88}~\shortcite{raz88}, thus resolving
the previously assumed ``contradiction'' between the experiments (see also
Sec.\ 3.1).
\subsection{Neutron capture on Gd--isotopes}
Gadolinium is one of six elements with two even s--only isotopes
($^{152}$Gd and $^{154}$Gd). Such isotopes are important in the detailed
investigation of the related s--process branchings which can help to
limit the physical conditions in the helium burning zones of Red Giants
(cf., Sect.~2.1). Neutron capture on $^{147}$Gd is also of
interest in theoretical studies concerning the so--called $\gamma$--process
in supernovae of type II, suggested by \citeauthor{Woos78a}~\shortcite{Woos78a},
and by \citeauthor{Rayet90}~\shortcite{Rayet90}.
In the case of Gd, the level density is high enough so that the direct--capture
contribution is negligible and calculations can be restricted to
the pure statistical model.

The statistical model (Hauser--Feshbach) is described in more
detail in Sect.~4.4.
For the statistical model
calculations presented in this section, we used the code SMOKER as described
in \citeauthor{Cowan91}~\shortcite{Cowan91}, but with an improved
level density description based
on \citeauthor{Ignat75}~\shortcite{Ignat75} and
\citeauthor{Ilji92}~\shortcite{Ilji92}, including thermal damping of
shell effects.
The level densities were the subject to the largest
uncertainties in the  most recent cross section calculations~\cite{Cowan91}.
The new description (\citeauthor{Rau95a}, \citeyear{Rau95a}, \citeyear{Rau95b},
\citeyear{Rau96})
reduces the number
of parameters compared to the best global fit as given in
\citeauthor{Cowan91}~\shortcite{Cowan91}
while lowering the mean deviation from experimental
data considerably (from a factor more than 3 down to a factor of about
1.5) and is expected to give a similar improvement for the
accuracy of level densities for nuclei far from stability.
Concerning masses, we made use of the most recent mass table~\cite{Audi93}
and the mass formula by \citeauthor{Moell95}~\shortcite{Moell95} where
experimental information was not available.

Recently, stellar neutron capture cross sections
for the Gd--isotopes with the mass numbers
$A$=152, 154, 155, 156, 157, and 158 have been determined experimentally
by \citeauthor{wis95}~\shortcite{wis95}. In order to test the approach
taken to calculate neutron capture on $^{148}$Gd
\cite{Rau95a}, we compared
our results to the experimental cross sections (see Table~\ref{crosstab}).
This experimental data has been obtained using the
total absorption detection employing a ball of scintillation crystals
as described in Sect.~3.1.2.

As can be seen, the values agree reasonably well, especially for odd
targets. It has to be emphasized, however, that no parameters of the
level density description were
adjusted especially to the Gd region, in order to preserve the reliable
predictive power for unstable isotopes from a {\em global} parameter fit.
Further work on improving the statistical model calculations is in
progress, and it is estimated that the theoretical cross sections will
get a global accuracy of about 30\% in future calculations.
\acknowledgements We thank the  Fonds zur F\"orderung der
wissenschaftlichen Forschung in \"Osterreich (project S7307--AST)
and the \"Osterreichische Nationalbank (project 5054)
for their support. TR is supported by an APART fellowship from the Austrian
Academy of Sciences.
One of us (HB) would like to thank F. Corvi
for valuable discussions. Part of this work was completed during a stay
at the Institute for Nuclear Theory, University of Washington, Seattle,
USA. HO and TR thank the INT for the support and hospitality.

\appendix

\section{Astrophysical reaction rates}
\label{astroapp}

The quantity most often quoted when dealing with nuclear reactions in
astrophysics is the nuclear reaction rate, measuring the number of
reactions per particle pair per second (it can also be generalized to
reactions involving more than two nuclei). It can be calculated from
the nuclear cross section $\sigma$ for a given reaction by folding it
with the velocity (i.e., energy) distribution of the particles involved.
In most astrophysical applications the nuclei are in a thermalized
plasma, yielding a Maxwell-Boltzmann velocity distribution.
The astrophysical reaction rate $R$ at a temperature $T$
can then be written as~\cite{Fowl75}
\begin{equation}
\label{RR}
R(T)=\left< \sigma v \right> = \left( \frac{8}{\pi m} \right)^{1/2}
\frac{1}{(k_{\rm B}T)^{3/2}} \int_0^{\infty} \sigma(E) E
\exp\left(-\frac{E}{k_{\rm B}T}\right) \, dE
\quad,
\end{equation}
with the reduced mass $m$ of the interacting particles and the
Boltzmann constant $k_{\rm B}$.

The threshold behavior of reaction cross sections is fundamental
in nuclear astrophysics because of the small projectile
energies in the thermonuclear region.
Near the threshold the reaction cross section can be written \cite{Bla62}
\begin{equation}
\label{CST}
\sigma_{if} = \frac{\pi}{k^2}\frac{-4kR{\rm Im}f_0}
{\left|f_0\right|^2}
\quad,
\end{equation}
where $f_\ell$ is the logarithmic derivative of the
scattering wave function for the $\ell$--th partial
wave at an appropriate nuclear radius $R$
\begin{equation}
\label{LD}
f_\ell = R \left(\frac{1}{u_\ell(r)}\frac{du_\ell(r)}{dr}\right)_{r=R}
= R \left(\frac{d \ln u_\ell(r)}{dr}\right)_{r=R}
\quad.
\end{equation}
Since the logarithmic derivative $f_0$ of the scattering wave
function is only weakly dependent on the projectile energy,
one obtains for low energies the well--known 1/v--behavior.

With increasing neutron energy higher partial waves with $\ell > 0$
contribute more significantly to the reaction cross section. Thus
the product $\sigma v$ becomes a slowly varying function of the
neutron velocity and one can expand this quantity into an
expansion in terms of $v$ or $\sqrt{E}$ around zero energy:
\begin{equation}
\label{AS}
\sigma v = S(0) + \dot{S}(0) \sqrt{E} + \ddot{S}(0) E
+ \ldots \quad .
\end{equation}
The quantity $S(E) = \sigma v$ is the astrophysical S--factor for
neutron--induced reactions and the dotted quantities represent
derivatives with respect to $E^{1/2}$. Notice that
the above astrophysical S--factor for neutron--induced
reactions is different from that for charged--particle induced
reactions. In the astrophysical S--factor for
charged--particle induced reactions also the penetration
factor through the Coulomb barrier has to be considered.

Inserting this into Eq.~\ref{RR}
we obtain for the reaction rate for neutron--induced reactions
\begin{equation}
\label{RRN}
\left<\sigma v\right> = S(0) + \left(\frac{4}{\pi}\right)^\frac{1}{2}
\dot{S}(0) (k_{\rm B}T)^\frac{1}{2} + \frac{3}{4} \ddot{S}(0) k_{\rm B}T + \ldots \quad .
\end{equation}

In most astrophysical neutron--induced reactions, neutron s--waves will
dominate, resulting in a cross section showing a 1/$v$--behavior
(i.e., $\sigma(E) \propto 1/\sqrt{E}$). In this case, the reaction
rate will become independent of temperature, $R={\rm const}$. Therefore
it will suffice to measure the cross section at one temperature
in order to calculate the rates for a wider range of temperatures.
Experiments using neutrons with a Maxwell-Boltzmann energy
distribution (of a temperature $T$)
directly measure the so--called Maxwellian Averaged Capture (MAC)
Cross Sections $\left< \sigma \right>_T$. The rate can then
be computed very easily by using
\begin{equation}
R=\left< \sigma v \right>=
\left< \sigma \right>_T v_T = {\rm const}\quad,
\end{equation}
with
\begin{equation}
v_T=\left(\frac{2kT}{m}\right)^{1/2}\quad.
\end{equation}

The mean lifetime $\tau_{\rm n}$
of a nucleus against neutron capture, i.e., the mean time between
subsequent neutron captures is inversely proportional to the available
number of neutrons $N_{\rm n}$ and the reaction rate $R_{\rm n\gamma}$:
\begin{equation}
\tau_{\rm n}=\frac{1}{N_{\rm n}R_{\rm n\gamma}}\quad.
\end{equation}
If this time is shorter than the beta--decay half--life of the nucleus,
it will be likely to capture a neutron before decaying. In this manner,
more and more neutrons can be captured to build up nuclei along an
isotopic chain until the beta--decay half--life of an isotope finally
becomes shorter than $\tau_{\rm n}$. With the very high neutron densities
encountered in several astrophysical scenarios,
isotopes very far--off stability can be synthesized.

\section{Expressions for cyclic activation}\label{A1}
The time constants for each cycle adjusted to the decay rate
$\lambda$ of the investigated isotope are the irradiation time
$t_{\rm b}$, the counting time
$t_{\rm c}$, the
waiting time $t_{\rm w}$ (the time to switch from the
irradiation to the counting phase),
and the total
time $T=t_{\rm b}+t_{\rm w}+t_{\rm c}+t'_{\rm w}$ ($t'_{\rm w}$ the time
to switch from the
counting to the irradiation phase).
The accumulated number of counts from a total of $n$ cycles,
$C=\sum_{i=1}^n C_i$, with the $C_i$,
the counts after the i--th cycle, irradiated by a neutron flux $\Phi_i$,
given by
\begin{eqnarray}
\label{eq1}
C & = & \epsilon_{\gamma}K_{\gamma}f_{\gamma}[1-\exp(-\lambda t_{\rm c})]
\frac{\exp(-\lambda t_{\rm w})}{1-\exp(-\lambda T)} N \sigma
[1-f'_{\rm b} \exp(-\lambda T)] \\\nonumber
&& \times \sum_{i=1}^n \int_0^{t_{\rm b}}\Phi_i \exp(-\lambda t)dt \quad ,
\end{eqnarray}
with
\begin{eqnarray}
f'_{\rm b}=\frac{\sum_{i=1}^n \int_0^{t_b}\Phi_i \exp(-\lambda t)
\exp[-(n-i)\lambda T]dt}{\sum_{i=1}^n \int_0^{t_b}\Phi_i \exp(-\lambda t)dt}
\quad .
\nonumber
\end{eqnarray}
The activities of nuclides with half--lives of several hours to days can also be
counted after
the end of the cyclic activation consisting of $n$ cycles:
\begin{eqnarray}
\label{eq2}
C_n & = & \epsilon_\gamma K_\gamma f_\gamma [1-\exp(-\lambda T_{\rm M})]\exp(-\lambda T_{\rm W})
N \sigma f'_b \\\nonumber
&& \times\sum_{i=1}^n \int_0^{t_{\rm b}}\Phi_i \exp(-\lambda t)dt \quad .
\end{eqnarray}
Here $T_{\rm M}$ is the measuring time of the Ge--detector and $T_{\rm W}$
the time elapsed between the
end of cyclic activation and beginning of the new data acquisition.

In the application of the cyclic activation method
the irradiation time $t_{\rm b}$ is chosen to be short compared to
the fluctuations of the neutron flux so that one can
integrate~\cite{bee94}:
\begin{eqnarray}
\int_0^{t_{\rm b}} \Phi_i \exp(-\lambda t)dt=
\lambda^{-1}[1-\exp(-\lambda t_{\rm b})]\Phi_i \quad . \nonumber
\end{eqnarray}
\section{Hauser--Feshbach formula}\label{A2}
The Hauser--Feshbach formula is given by
\begin{eqnarray}
\sigma^{\mu \nu}_{i} (j,o;E) & = &
\frac{\pi \hbar^2 }{(2 \mu_{\alpha} E) (2J^\mu_i+1)(2J_j+1)}
\\\nonumber
&& \times \sum_{J,\pi} (2J+1)\frac{T^\mu_j (E,J,\pi ,E^\mu_i,J^\mu_i,
\pi^\mu_i) T^\nu_o (E,J,\pi,E^\nu_m,J^\nu_m,\pi^\nu_m)}
{T_{\rm tot} (E,J,\pi)}
\label{hf}
\end{eqnarray}
for the reaction $i^\mu (j,o) m^\nu$ from the target
state $i^{\mu}$ to the exited state $m^{\nu}$ of the final nucleus, with
center of mass energy $E$ and reduced mass $\mu$. The
angular momentum quantum number $J$ denotes the
spin, $E$ the excitation energy, and $\pi$ the parity of excited states.
When these properties are used  without subscripts they describe the
compound
nucleus, subscripts refer to states of the participating nuclei in the
reaction $i^\mu (j,o) m^\nu$
and superscripts indicate the specific excited states.
Experiments measure $\sum_{\nu} \sigma_{i} ^{0\nu} (j,o;E)$,
summed over all excited states of
the final nucleus, with the target in the ground state. Target states
$\mu$ in
an astrophysical plasma are thermally populated and the astrophysical
cross
section $\sigma^*_{i}(j,o)$ is given by
\begin{equation}
\label{thermcs}
\sigma^*_{i} (j,o;E) = \frac{\sum_\mu (2J^\mu_i+1)
\exp\left(-\frac{E^{\mu}_i}{k_{\rm B}T}\right)
\sum_\nu \sigma^{\mu \nu}_{i}(j,o;E)}{\sum_\mu (2J^\mu_i+1)
\exp\left(-\frac{E^{\mu}_i}{k_{\rm B}T}\right)} \quad .
\end{equation}
The summation over $\nu$ replaces $T_o^{\nu}(E,J,\pi)$ in Eq.~\ref{hf} by
the total transmission coefficient
\begin{eqnarray}
T_o (E,J,\pi) & = & \sum^{\nu_m}_{\nu =0}
T^\nu_o(E,J,\pi,E^\nu_m,J^\nu_m, \pi^\nu_m) \\\nonumber
 & & + \int\limits_{E^{\nu_m}_m}^{E-S_{m,o}} \sum_{J_m,\pi_m}
T_o\left(E,J,\pi,E_m,J_m,\pi_m\right)
\rho\left(E_m,J_m,\pi_m\right) dE_m  \quad .
\label{trans}
\end{eqnarray}
Here $S_{m,o}$ is the channel separation energy, and the summation over
excited
states above the highest experimentally
known state $\nu_m$ is changed to an integration over the level density
$\rho$.
The summation over target states $\mu$ in Eq.~\ref{thermcs}
has to be generalized
accordingly.
\section{Expressions for DC and DWBA}\label{A3}
The differential cross section for the transfer reaction
$a+A \rightarrow b+B$ with $a-x=b$, $A+x=B$ (stripping) using light
projectiles and ejectiles (for $a \leq 4$ and $x=1$ or $x=3$) is given in
zero--range DWBA
by \cite{Sat83,Gle83}
\begin{equation}
\label{DWBA}
\frac{d\sigma}{d\Omega} = \frac{\mu_\alpha\mu_\beta}
{(2\pi \hbar^{2})^{2}}\frac{k_\beta}{k_\alpha}
\frac{2I_{B}+1}{2I_{A}+1} \sum_{\ell sj}C^{2}{\cal S}_{\ell j}N
\frac{\sigma_{\ell sj}(\vartheta)}{2s+1}
\end{equation}
with the zero--range normalization constant
\begin{equation}
\label{Norm}
N=\frac{1}{2}a D_0^2 \quad .
\end{equation}
The reduced cross section without spin--orbit coupling is
\begin{equation}
\label{RCS}
\sigma_{\ell sj}(\vartheta)=\sum_{m}\left|t_{\ell sj}^{m}\right|^{2}
\end{equation}
with the reduced transition amplitude
\begin{equation}
t_{lsj}^m=\frac{1}{2\ell +1}\int \; d{\bf r}
\chi_\beta^{(-)*}\left({\bf k}_{\beta},{A}{B} {\bf r}\right)
u_{\ell j}(r) [i^{\ell}Y_\ell ^{m}({\bf \hat r})]^{*}
\chi_\alpha^{(+)}({\bf k}_{\alpha},{\bf r}) \quad.
\end{equation}
The quantities $\mu_{\alpha}$, $\mu_{\beta}$ and $k_{\alpha}$,
$k_{\beta}$ are the reduced masses and wave numbers in the entrance
channel $\alpha$ and exit channel $\beta$, respectively. The
spin and magnetic spin quantum numbers of the
projectile, ejectile, target and residual nucleus are given by
($I_{\rm a}$, $I_{\rm b}$, $I_{\rm A}$, $I_{\rm B}$)
and ($M_{\rm a}$, $M_{\rm b}$, $M_{\rm A}$, $M_{\rm B}$), respectively.
The orbital angular momentum quantum number $\ell$,
the spin quantum number $s$ and the total angular momentum quantum number
$j$ refer to the cluster $x$ bound in the residual nucleus $B$. The
optical wave functions in the entrance and exit channels are
characterized by $\chi^{(+)}$ and the time--reversed solution
$\chi^{(-)}$. The spectroscopic factor and the isospin
Clebsch--Gordan coefficient for the partition $B=A+x$ are given
by $C$ and ${\cal S}_{\ell j}$, respectively.
Expressions similar to the above equations will be obtained
if the finite range of the interaction potential is taken into account
\cite{Sat83}.

The above expressions are written for a stripping reaction, where
the nucleon
cluster $x$ is stripped from the projectile $a$. The
corresponding formulae for a pick--up reaction can be obtained easily
with the help of the reciprocity theorem for the
reduced cross sections \cite{Sat83}.

The direct capture
process $a+A \rightarrow \gamma+B$, which is entirely
electromagnetic, is treated
in first--order perturbation theory. As examples, we quote the expressions
for the electric dipole E1 capture \cite{Chr61,Rol73}:
\begin{eqnarray}
\label{DC}
\sigma_{\rm E1} & = & \frac{16\pi}{9}
\left(\frac{E_\gamma}{\hbar c}\right)^3
\frac{e^2\mu_\alpha^3}{\hbar^2k_\alpha}
\frac{3}{\left(2I_{\rm a}+1\right)\left(2I_{\rm A}+1\right)}
\left(\frac{Z_{\rm a}}{m_{\rm a}} -\frac{Z_{\rm A}}{m_{\rm A}}\right)^2
C^{2}{\cal S}_{\ell_ \beta J_\beta}\\\nonumber
&& \times \sum_{\ell_ \alpha J_\alpha}
\left(2J_\beta+1\right)\left(2J_\alpha+1\right)
\max\left(\ell_\alpha,\ell_\beta\right)\\\nonumber
&&\times \left\{\begin{array}{ccc}1 & \ell_{\beta} & \ell_{\alpha}\\
I & J_{\alpha} & J_{\beta}\end{array}\right\}^2
a_{I}^2 \left\vert R_{1 \beta \alpha} \right\vert^2
\end{eqnarray}
with the radial integral
\begin{equation}
R_{1 \beta \alpha}=\frac{1}{k_{\alpha}}
\int \;dr
u_{\beta}^{*}(r){\cal O}_{E1}(r)\chi_{\alpha}(r) \quad .
\end{equation}
The coefficients $a_{I}^2$ are calculated in LS coupling to
\begin{eqnarray}
a_I^2 & = & (2I+1)\left(2I_{\rm A}+1\right)
\left(2L_{\rm B}+1\right)\left(2S_{\rm B}+1\right)\\\nonumber
&&\times \left\{\begin{array}{ccc}I & L_{\rm A} & S_{\rm B} \cr
L_{\rm B} & I_{\rm B} & \ell_{\beta}\end{array}\right\}^2
\left\{\begin{array}{ccc}I & L_{A} & S_{B}\\
I_{\rm A} & I_{\rm a} & I_{\rm A}\end{array}\right\}^2 \quad.
\end{eqnarray}
In the above expressions, the energy of the emitted photon
is $E_{\gamma}$. The charge and mass of the projectile
and target nucleus are $Z_{\rm a}$, $m_{\rm a}$, $Z_{\rm A}$ and
$m_{\rm A}$, respectively. The orbital
and total angular momentum quantum numbers of the nuclei
in the entrance and exit channels are $\ell_{\alpha}$,
$J_{\alpha}$, $\ell_{\beta}$ and $J_{\beta}$, respectively.
The spin quantum number, the orbital
and total angular momentum quantum numbers are characterized by
$S$, $L$ and $I$, respectively, with indices $a$, $A$ and $B$
corresponding to the projectile, target and
residual nucleus, respectively. The symbol
$\left\{\matrix{\ldots}\right\}$
is the $6j$ symbol. The radial wave functions in the entrance and
exit channels are given by $\chi_{\alpha}$ and $u_{\beta}$,
respectively. The spectroscopic factor and the isospin
Clebsch--Gordan coefficient for the partition $B=A+a$ are given
by $C$ and ${\cal S}_{\ell_{\beta} J_{\beta}}$, respectively.
The ${\cal O}_{{\rm E}\ell}$ are the multipole
operators.

\end{document}